\documentclass[twocolumn]{raa_twocolumn}            % referee version: for submission

\usepackage{graphicx,times}             %for PS/EPS graphics inclusion, new
\usepackage{natbib}
\usepackage{amssymb,amsmath}
\bibpunct{(}{)}{;}{a}{}{,}

\usepackage[pagebackref=true]{hyperref}

\begin{document} 

\title{MIU\textsuperscript{2}Net: weak-lensing mass inversion using deep learning with nested U-structures\thanks{All python code and data that we used to generate the figures in the paper are available at \href{https://github.com/MIU2NET/miu2net}{https://github.com/MIU2NET/miu2net}.}}

\author{
          Han W.G.\inst{1}
          \and
          An Zhao\inst{2}
          \and
          Xingyue Chen\inst{3}
          \and
          Ran Li\inst{1}
            \and
          Rui Li\inst{4}
          \and
          Xiangkun Liu\inst{2}
          \and 
          Zhao Chen \inst{5}
          \and
          Yu Yu \inst{5}
          }

   \institute{School of Physics and Astronomy, Beijing Normal University,  Beijing 100875, China
         \and
            South-Western Institute for Astronomy Research, Yunnan University, Kunming 650500, China
         \and
            School of Astronomy and Space Science, University of Chinese Academy of Sciences, Beijing 100049, China
         \and
            Institute for Astrophysics, School of Physics, Zhengzhou University, Zhengzhou, 450001, China
        \and
            Department of Astronomy, School of Physics and Astronomy, Shanghai Jiao Tong University, Shanghai 200240, China
             }

\date{Received XXX; accepted XXX}

\abstract{
One of the primary goals of next-generation gravitational lensing surveys is to measure the large-scale distribution of dark matter, which requires accurate mass inversion to convert weak-lensing shear maps into convergence ($\kappa$) fields. This work develops a mass inversion method tailored for upcoming space missions such as CSST and \textit{Euclid}, aiming to recover both the mass distribution and the convergence power spectrum with high fidelity. We introduce MIU$^{2}$Net, a versatile deep-learning framework for $\kappa$-map reconstruction based on the U$^{2}$-Net architecture. A new loss function is constructed to jointly estimate the convergence field and its frequency-domain energy distribution, effectively balancing optimal mean squared error and optimal power-spectrum recovery. The method incorporates realistic observational effects into shear fields, including shape noise, reduced shear, and complex masks. Under noise levels anticipated for future space-based lensing surveys, MIU$^{2}$Net recovers the convergence power spectrum with 4\% uncertainties up to $l \simeq 500$, significantly outperforming Wiener filtering and MCALens. Beyond two-point statistics, the method accurately reconstructs the convergence distribution, peak centroid, and peak amplitude. Compared to other learning-based approaches such as DeepMass, MIU$^{2}$Net reduces the root-mean-square error by 5\% without smoothing and by 38\% with a 1-arcmin smoothing scale. MIU$^{2}$Net represents a substantial advancement in mass inversion methodology, offering improved accuracy in both RMSE and power-spectrum reconstruction. It provides a promising tool for mapping dark matter environments and large-scale structures in the era of next-generation space lensing surveys.
}

\maketitle

\keywords{gravitational lensing: weak -- methods: statistical -- dark matter -- large-scale structure of Universe}

\section{Introduction}\label{sec:Introduction}

Gravitational lensing is the relativistic bending of light as it travels to us. The degree of bending depends on the gravitational potential of intervening mass, as well as the geometry of the lensing system. In the weak lensing regime, we study the distortions in background galaxy shapes to infer the foreground mass distribution through which light travels. Weak lensing information can be accessed by two equivalent fields: the (reduced) shear field\footnote{For a detailed discussion of the shear and the reduced shear, please refer to the end of Sect.~\ref{sec:Formalism}.} \(\gamma_{1,2}\) and the convergence field \(\kappa\). The (reduced) shear field is an observable that describes the stretching of galaxy shapes. The convergence field, however, is not an observable, but it is directly linked to the projected matter density. The convergence field has extensive applications in  large-scale structure identification \citep[e.g.][]{Christopher2018id,Mead2010id,Tam2020id}, constraining the cosmological model \citep[e.g.][]{ Liu2015peakcount,Davies2021cos,LiuJia2019neutrino}, studies of non-Gaussianity in the universe \citep[e.g.][]{Pires2020KS+}, and tests of gravity \citep[e.g.][]{Liu2016grav, Ling2015grav}. Therefore, a transformation from \(\gamma_{1,2}\) field to \(\kappa\) field is much desired. This transformation is typically referred to as a ``mass inversion'' because we are recovering mass distribution from observed shear.

%The convergence field also compresses complex shear into a scalar field, making calculations much less computationally expensive. 

However, mass inversion is an ill-posed inverse problem due to an excessive amount of noise. It is well known that the intrinsic shape dispersion of galaxies (known as ``shape noise,'' see Sect.~\ref{sec:Shape Noise}) is much more significant than the typical distortion from weak lensing. Moreover, the observed shear map is pixelated and sampled at discrete locations; it also has empty regions because of survey geometry, bright star masks, CCD defects, etc. We also cannot know the mean convergence due to a fundamental limit called the mass-sheet degeneracy \citep{Falco1985MassSheet, SchneiderSeitz1995-I, BradacSchneider2004MassSheet}. These observational limitations further complicate mass inversion. Starting with \citet{KS93}, many mass inversion techniques have been proposed, including Kaiser-Squires (KS) deconvolution and its variants \citep{KS93, Pires2009KS+, Deriaz2012KS+, Pires2020KS+}, Wiener filtering (Gaussian prior) \citep{Lahav1994WF, Zaroubi1995WF}, sparsity prior \citep{Lanusse2016Glimpse, Jeffrey2018sparse, Price2021Glimpse}, and Bayesian methods with other physically motivated priors \citep{Jasche2013bayesian, Porqueres2021bayesian, Fiedorowicz2022karmma}, all assuming mean convergence of zero. {\color{black}Additionally, \citet{Starck2021MCALens} proposed an iterative approach (MCALens) by combining Wiener filtering with a sparsity-based non-Gaussian component.} With the rapid rise of machine learning, mass inversion based on deep learning techniques started to flourish. \citet{Jeffrey2020ML} pioneered this interdisciplinary work by designing {\color{black} DeepMass,} a UNet convolutional neural network (CNN) for DES mass inversion. \citet{Hong2021ML} applied the CNN architecture on galaxy clusters, and \citet{Shirasaki2019ML} used variants of generative adversarial network (GAN) to denoise convergence maps. However, despite the visual competitiveness of deep learning reconstructions, all proposed CNN models minimize only a pixel-wise loss function, thereby limiting the physical significance of CNN-based convergence maps. GAN variants are different in that they include learned networks as parts of the loss function, but they too only optimize in the spatial domain. Specifically, \citet{Shirasaki2019ML} show that the classical conditional GAN model (\href{https://github.com/phillipi/pix2pix}{pix2pix}; \citealt{pix2pix}) has limited capacity in Fourier space, yielding a sub-optimal estimate of the convergence power spectrum.

% Additionally, mass inversion is troubled by a fundamental limit called the mass-sheet degeneracy, where the convergence \(\kappa\) field can only be determined up to the degeneracy transformation \(\kappa \rightarrow \kappa_\lambda = \lambda\kappa + (1-\lambda)\), where \(\lambda\) is an arbitrary constant \citep{Falco1985MassSheet, SchneiderSeitz1995-I, BradacSchneider2004MassSheet}. This means that a rescaling of the ``true'' convergence field, combined with the addition of a homogeneous surface mass density \(1-\lambda\) , is an equally good solution as \(\kappa\). Many methods have been proposed to lift the mass-sheet degeneracy \citep{SeitzSchneider1997-III, BradacSchneider2004MassSheet, Rexroth2016Moments, Cremonese2021GW}, but all methods require additional data or assumptions (e.g. accurate redshift distribution, flexion fields, gravitational wave inferences, etc.). For undercritical lenses, it has more or less been proven impractical to break the mass-sheet degeneracy without using weakly-constrained information such as the magnification effect \citep{Broadhurst1995Magnification, SeitzSchneider1997-III}.

\begin{figure*}
    \centering
    \includegraphics[width=1\linewidth]{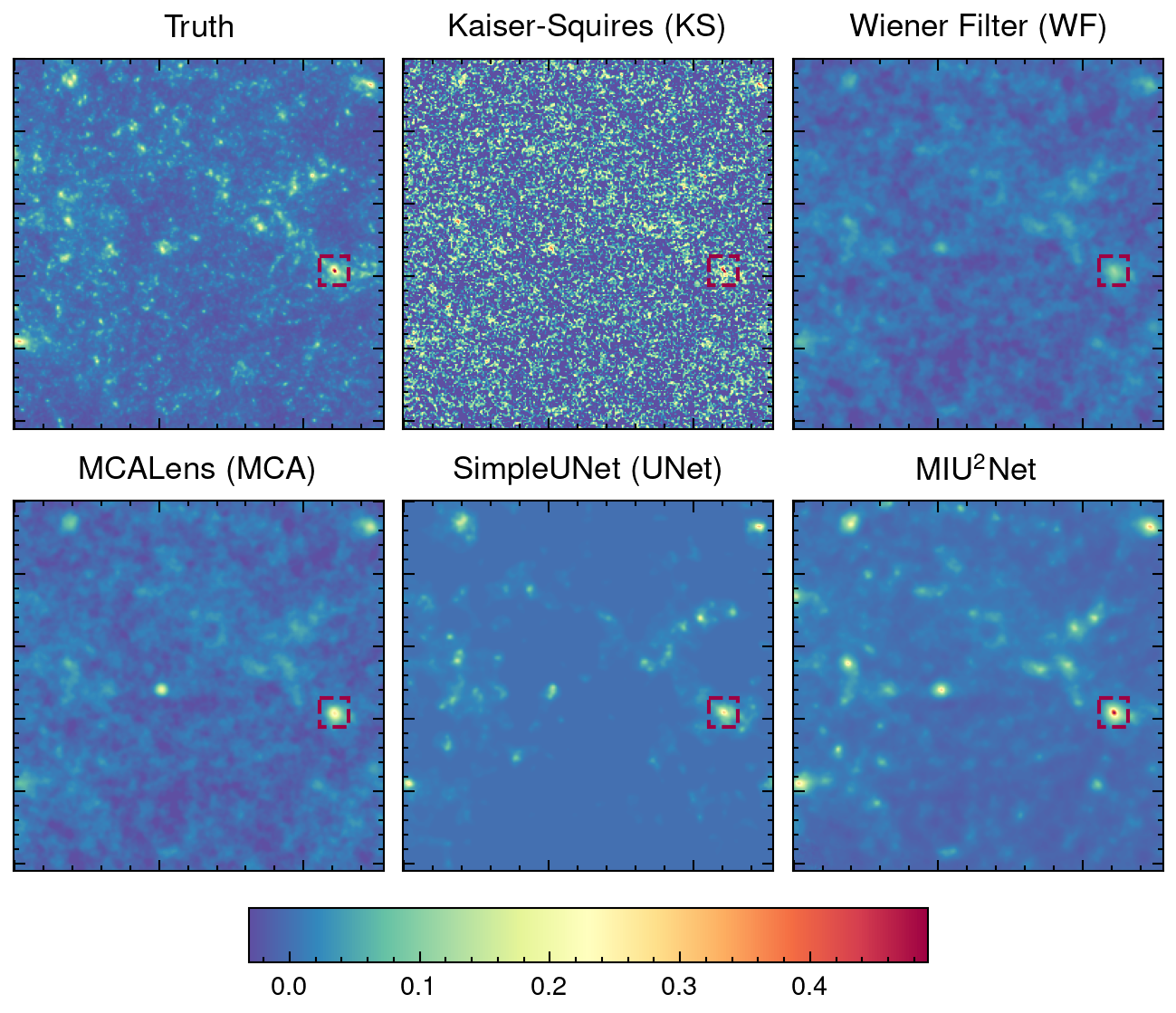}
    \caption{Visual comparison between Truth, our method (MIU\textsuperscript{2}Net), and established methods including Kaiser-Squires (KS), Wiener filtering (WF), MCALens (MCA), and {\color{black}SimpleUNet (UNet, based on DeepMass)}. Each panel covers a \(1.75 \times 1.75\) $\deg^2$ Euclidean patch of sky. The shear maps used for reconstruction have shape noise corresponding to galaxy number density \(n_g = 20 \, \rm{arcmin}^{-2}\). We visualize in detail the scarlet dashed square region in each panel in Fig.~\ref{fig:profile3d}.}
    \label{fig:visual}
\end{figure*}

Building on previous works \citep{Jeffrey2020ML,Hong2021ML,Shirasaki2019ML}, we propose a {\color{black}versatile} deep learning framework for mass inversion (MIU\textsuperscript{2}Net) based on a nested U-shaped CNN. We construct a new loss function to jointly estimate the convergence field and its frequency domain energy distribution, seeking a balance between optimal MSE and optimal power spectrum. Under a noise level achievable by upcoming space lensing surveys, we find that MIU\textsuperscript{2}Net recovers the convergence power spectrum with \(4\%\) uncertainties up to \(l \simeq 500\), \(1060\%\) more accurate than Wiener filtering {\color{black}and $930\%$ more accurate than MCALens} with optimal power spectrum. MIU\textsuperscript{2}Net can also recover the {\color{black}convergence distribution, peak centroid, and peak abundance.} A visual comparison between MIU\textsuperscript{2}Net, Kaiser-Squires deconvolution (KS), Wiener filtering (WF), MCALens (MCA), and {\color{black}SimpleUNet (UNet, based on DeepMass)} is shown in Fig.~\ref{fig:visual}. We introduce the details of the method and results in this paper.

The structure of this paper is as follows. In Sect.~\ref{sec:Weak Gravitational Lensing}, we introduce the basic formalism of weak gravitational lensing as well as the major challenges in mass inversion such as shape noise, mass-sheet degeneracy, and missing data. In Sect.~\ref{sec:Mass Inversion Methods}, we briefly outline the current mass inversion methods including KS, WF, and {\color{black} UNet}. We introduce our method, MIU\textsuperscript{2}Net, in Sect.~\ref{sec:MIU2Net}, detailing our training data, network architecture, and loss function. We demonstrate the reconstruction results in Sect.~\ref{sec:Results} and compare the MIU\textsuperscript{2}Net results to {\color{black}unsupervised methods such as KS, WF, MCA, and supervised deep learning methods such as {\color{black} UNet}}.

\section{Weak Gravitational Lensing}\label{sec:Weak Gravitational Lensing}

\subsection{Formalism}\label{sec:Formalism}

In this section, we introduce the basics of weak lensing formalism and the observational challenges it faces. (For a detailed review of weak lensing formalism, please see \citealt{BartelmannSchneider2001Review, Mandelbaum2018Review}.) Consider the 2D lensing potential \(\psi(\bm\theta)\) that encodes the deflection of light rays. The reduced (scaled) deflection angle \(\bm{\alpha}\) can be written as the gradient of the potential
\begin{equation}
    \bm{\alpha} = \nabla \psi,
\end{equation}
and the convergence \(\kappa\) is related to \(\psi\) through the 2D Poisson equation
\begin{equation}
    \kappa = \frac{1}{2} \left( \partial_1^2 + \partial_2^2 \right) \psi = \frac{1}{2}\nabla^2 \psi,
    \label{pois}
\end{equation}
where the subscripted partial derivatives \(\partial_1\), \(\partial_2\) relates to the two angular coordinates \(\theta_1\), \(\theta_2\) on the celestial sphere. The two-component shear is expressed as a complex field, where each component is described by \(\psi\) as
\begin{equation}
    \gamma_1 = \frac{1}{2} \left( \partial_1^2 - \partial_2^2 \right)\psi,
    \quad
    \gamma_2 = \partial_1\partial_2\psi,
\end{equation}
and the complex shear field
\begin{equation}
    \gamma = \gamma_1 + \mathrm{i}\gamma_2 = |\gamma| \mathrm{e}^{2\mathrm{i}\varphi}.
\end{equation}
Note that all lensing quantities can be derived from the lensing potential \(\psi\), thereby setting the relations between them. Through the lensing potential, we can express complex shear in terms of a convolution on the convergence, or
\begin{equation}
    \gamma(\bm\theta) = \frac{1}{\pi} \int_{\mathbb{R}^2}
    {\rm{d}^2 \theta' \, \mathcal{D}(\bm\theta-\bm\theta')} \,
    \kappa(\bm\theta'),
    \label{conv}
\end{equation}
where \(\mathcal{D}(\bm\theta) = -1 / (\theta_1 - \mathrm{i}\theta_2)^2\). This equation can be discretized into a linear model using matrix notation
\begin{equation}
    \bm\gamma = \mathbf A \bm\kappa, \label{linear}
\end{equation}
where \(\bm\gamma\) and \(\bm\kappa\) are vector quantities for shear and convergence sampled on an image grid. Throughout the paper, we assume that the image patch size is small enough for sky geometry to be Euclidean.

In the weak lensing regime, the source galaxies are much smaller than the angular scale where lensing properties vary. This allows us to use Born approximation and write out the amplification matrix \(\mathcal{A}\), a Jacobian that describes the lensing transformation from the source plane to the image plane
\begin{equation}
    \mathcal{A}(\bm\theta) = 
    \begin{pmatrix}
	1 - \kappa - \gamma_1 & -\gamma_2 \\
	-\gamma_2 & 1 - \kappa + \gamma_1
    \end{pmatrix}.
\label{jacob}
\end{equation}
Rewriting the amplification matrix and factoring out \((1-\kappa)\), we have
\begin{equation}\label{eq:reduced-shear}
    \mathcal{A}(\bm\theta) = \left( 1 - \kappa \right)
    \begin{pmatrix}
	1 - g_1 & -g_2 \\
	-g_2 & 1 + g_1
    \end{pmatrix},
    \quad g_{1,2} = \frac{\gamma_{1,2}}{1-\kappa},
\end{equation}
where \(g = g_1 + \mathrm{i}g_2\) is the reduced shear field and is a direct observable. In practice, \(\bm\gamma, \bm\kappa \ll 1\) for most weak lensing regions, so that the reduced shear and the true shear are approximately the same, or \(g \simeq \gamma\). This allows us to extend the linear relation in \((\ref{linear})\) for reduced shear. However, for regions near convergence peaks, \(\bm\kappa\) is not negligible at small angler scales \citep{White2005SmallAngle, Shapiro2009SmallAngle}, creating a significant discrepancy between shear and reduced shear. It is therefore preferable to reconstruct \(\bm\kappa\) from the reduced shear field \(\bm g\). {\color{black}For fair comparison, we use the shear field instead of the reduced shear field for all reconstruction methods. In practice, our method can easily incorporate reduced shear by augmenting the training set via (\ref{eq:reduced-shear}).}

\subsection{Shape Noise}\label{sec:Shape Noise}

Weak lensing information is only sampled at discrete locations where galaxy ellipticities are measured. However, galaxies are not intrinsically circular, so the observed ellipticity is made up of both the galaxy's intrinsic ellipticity and the distortion from weak lensing. This results in a ``shape noise'' order of magnitudes larger than the weak lensing signal. As shown by \citet{Waerbeke2000noise}, the observed shear \(\gamma_{\rm{obs}}\) can be modelled as true shear plus an addictive shape noise \(\epsilon\), or
\begin{equation}
    \gamma_{\rm{obs}} \approx g+ \epsilon \approx \gamma + \epsilon.
\end{equation}
In matrix notation, we write the observed shear as
\begin{equation}
    \bm\gamma_{\rm{obs}} = \bm\gamma + \mathbf n = \mathbf{A} \bm\kappa + \mathbf{n},
    \label{prob}
\end{equation}
where \(\mathbf{n}\) is a discrete Gaussian random field with zero mean and variance
\begin{equation}
    \sigma_n^2 = \frac{\sigma_\epsilon^2}{2}
						 \frac{1}{\theta_s^2 \, n_g}, \label{noise}
\end{equation}
where \(\sigma_\epsilon\) is the root-mean-square (RMS) amplitude of the galaxy intrinsic ellipticity distribution, \(\theta_s\) is the pixel side length in arcmin, and \(n_g\) is the galaxy number density per pixel. Throughout the paper, we adopt \(\sigma_\epsilon = 0.4\) and galaxy number density of $20 \, \rm{arcmin}^{-2}$, which are conservative values for next-generation ground- and space-based surveys \citep{Mandelbaum2018noise}. We can then draw a noise realization \(\mathbf n\) with the same shape as \(\bm\gamma\) and model shape noise by simple matrix addition. For an impression of scale, we have the RMS amplitude of noise about \(19\) times larger than that of the signal.

\subsection{Mass-Sheet Degeneracy}\label{sec:Mass-Sheet Degeneracy}

A fundamental limit of weak lensing mass mapping is the mass-sheet degeneracy. (For a detailed discussion, please refer to \citealt{Falco1985MassSheet} in the strong lensing regime, and \citealt{SchneiderSeitz1995-I, BradacSchneider2004MassSheet} for weak lensing.) Consider the transformation of the lensing potential
\begin{equation}
    \psi(\bm\theta) \rightarrow \psi_\lambda(\bm\theta) = 
    \frac{1-\lambda}{2} |\bm\theta|^2 + \lambda\psi(\bm\theta),
\end{equation}
so that we satisfy the Poisson equation~\((\ref{pois})\). \(\lambda\) here is an arbitrary constant. It then follows that the Jacobian in \((\ref{jacob})\) behaves as
\(\mathcal{A}_\lambda(\bm\theta) = \lambda \mathcal{A}(\bm\theta)\), and the convergence as
\begin{equation}
    \kappa(\bm\theta) \rightarrow \kappa_\lambda(\bm\theta) = 
    \lambda\kappa(\bm\theta) + (1-\lambda). \label{eq:mass-sheet}
\end{equation}
The first term is a linear rescaling of the ``true'' convergence field \(\kappa(\bm\theta)\), and the second term corresponds to the addition of a homogeneous surface mass density \(\kappa_0 = 1 - \lambda\). The convergence \(\kappa\) field can thus only be determined up to this degeneracy transformation. The mass-sheet degeneracy fundamentally limits weak lensing applications, because the reconstructed convergence field can only probe the \emph{fluctuations} of projected matter density on top of an unknown mean, instead of the \emph{absolute} projected matter density. This degeneracy cannot be lifted using ellipticities of background sources alone \citep{SeitzSchneider1997-III} and fundamentally limits weak lensing in determining the absolute mass distribution.

\subsection{Missing Data}\label{sec:Missing Data}

Another strong observational limitation is missing data. This is due to complex survey geometry, masking of bright stars, image reduction masks, defects in CCD sensors, etc. The observed shear tensor is therefore further modified by a binary mask operator specific to each image. Missing data is usually filled with \(0\), or, more sophisticatedly, filled by inpainting methods using Discrete Cosine Transform (DCT) \citep{Pires2009KS+, Pires2020KS+}. We approach the missing data problem in a similar vein as DCT, guiding deep learning to learn a frequency domain energy distribution inside the masked area.

\section{Mass Inversion Methods}\label{sec:Mass Inversion Methods}

In this section, we briefly introduce some of the standard methods of weak lensing mass inversion, then go on to illustrate the deep learning approach. We will then build on the standard deep learning framework and introduce MIU\textsuperscript{2}Net in the next section.

\subsection{Kaiser-Squires (KS)}\label{sec:Kaiser-Squires (KS)}

\citet{KS93} have shown that we can reconstruct the convergence field in the Fourier domain. Consider Eq.~\((\ref{conv})\) where the complex shear \(\gamma\) is a convolution on \(\kappa\) with the kernel \(\mathcal{D}\). We know that a convolution in the spatial domain is equivalent to an element-wise multiplication in the Fourier domain, therefore
\begin{equation}
    \tilde{\gamma}(\bm k) = \pi^{-1} \tilde{\mathcal{D}}(\bm k) \tilde{\kappa}(\bm k),
    \quad \bm k \neq 0 \label{fourier},
\end{equation}
where \(\bm k\) is the wavevector, and the tildes denote Fourier transforms. The Fourier transform of \(\mathcal{D}(\bm\theta)\) is
\begin{equation}
    \tilde{\mathcal{D}}(\bm k) = \pi \frac{\left( k_1^2 - k_2^2 + 2 \mathrm{i} k_1 k_2 \right)}{|\bm k|^2},
\end{equation}
which satisfies \(\tilde{\mathcal{D}}(\bm k) \tilde{\mathcal{D}}^*(\bm k) = \pi^2\). Inverting Eq.~\((\ref{fourier})\) then yields
\begin{equation}
    \tilde{\kappa}(\bm k) = \pi^{-1} \tilde{\gamma}(\bm k) \tilde{\mathcal{D}}^*(\bm k),
    \quad \bm k \neq 0.
\end{equation}
Transforming back to the spatial domain, we obtain the mass inversion
\begin{equation}
    \kappa(\bm\theta) - \kappa_0 = \frac{1}{\pi} 
    \int_{\mathbb{R}^2}{\rm{d}^2 \theta' \, 
    \mathcal{D}^*(\bm\theta-\bm\theta')} \,
    \gamma(\bm \theta').
\end{equation}
The arbitrary constant \(\kappa_0\) arises because the inversion is undefined when \(\bm k = 0\), which is the effect of the mass-sheet degeneracy. In practice, it is often assumed that the mean convergence vanishes for a large enough field, i.e., \(\kappa_0 = 0\). In addition, the standard KS deconvolution does not take into account noise, reduced shear, or masked data, and can therefore introduce errors in the reconstruction.

\subsection{Wiener Filtering (WF)}\label{sec:Wiener Filtering (WF)}

Another widely used mass inversion method is Wiener filtering \citep{Lahav1994WF, Zaroubi1995WF}. The Wiener filter is the {\color{black} linear minimum-variance solution for the general recovery problem in Eq.~\((\ref{prob})\), assuming a Gaussian prior.} If the noise is uncorrelated, Wiener filtering reconstructs convergence map \(\bm{\hat{\kappa}}_\mathbf{W}\) by the linear relation
\begin{equation}
    \bm{\hat{\kappa}}_\mathbf{W} = \mathbf{W} \bm\gamma_\mathrm{obs}, \label{wf}
\end{equation}
where \(\mathbf{W}\) is the Wiener filter
\begin{equation}
    \mathbf{W} = \left( \mathbf{A \Sigma_s A^\dagger +\Sigma_n} \right)^{-1} 
    \mathbf{A^\dagger \, \Sigma_s}.
\end{equation}
Here, \(\mathbf{\Sigma_s} = \langle \bm\kappa \bm\kappa^\dagger \rangle\) and \(\mathbf{\Sigma_n} = \langle \mathbf{nn}^\dagger \rangle\) are the covariance matrices for the convergence signal and noise, respectively. \(\mathbf{A}\) is the linear model matrix in Eq.~(\ref{linear}), and \(\mathbf{A^\dagger}\) is its conjugate transpose. From the Bayesian perspective, WF assumes a zero-mean Gaussian prior in the form
\begin{equation}
    p(\bm\kappa | \mathbf{\Sigma_s}) = 
    |2\pi \mathbf{\Sigma_s}|^{-\frac{1}{2}} \,
    \exp \left[ -\frac{1}{2} \bm\kappa^\dagger \, \mathbf{\Sigma_s} \, \bm\kappa \right],
\end{equation}
and the reconstruction result \(\bm{\hat{\kappa}}_\mathbf{W}\) is the \emph{maximum a posteriori} (MAP) solution of \(p(\bm\kappa | \bm\gamma)\), assuming that both signal and noise are gaussian random fields. If \(\mathbf{\Sigma_n} = 0\), WF reduces to KS deconvolution \citep{Simon2009WF-KS}. The Gaussian prior is physically motivated on large scales, but it is less appropriate for smaller scales where structure formation is non-linear. Like KS, WF assumes a zero-mean convergence during implementation.

\citet{Jeffrey2021AllMethods} point out that the MAP solution \(\bm{\hat\kappa}\) that maximizes \(p(\bm{\kappa} | \bm{\gamma})\) is not necessarily the MAP solution for the transformed probability \(p(f(\bm{\kappa}) | \bm\gamma)\). WF is thus not ideal for inferring cosmology, because a statistic on the Wiener posterior is no longer the most probable statistic. To estimate a given statistic \(\mu\), we can instead sample from the joint posterior \(p(\bm\kappa, \mu | \bm\gamma)\). Several works have shown that a joint estimate of the convergence map and its power spectrum is possible \citep{Wandelt2004WFps, Alsing2017WFps}.

\subsection{Deep Learning}\label{sec:Deep Learning}

The burst of deep learning in recent years has brought paradigm shifts to many academic fields. Compared to task-specific solvers, a self-learned gigantic parametric model seems to perform much better on ill-posed inverse problems. The model is often intentionally vague and non-transparent, yet it generalizes very well and handles unseen data with surprising sophistication. In the language of deep learning, we can approach the inverse problem in Eq.~\((\ref{prob})\) as a supervised image-to-image translation problem with these general steps:
\begin{enumerate}
\item
  Prepare a large ``clean'' dataset on the order of thousands or even millions of images. For weak lensing mass inversion, this means obtaining many shear and convergence realization pairs from simulations. This paired dataset will be the source from which the model learns a transformation.
\item
  Degrade the images-to-be-learned by noise realizations, random data masks, reduced shear, etc., so that the shear images resemble real observations.
\item
  Construct a parametric model to learn the transformation from noisy (reduced) shear to true convergence. The model represents a Universal Function Approximator, where it essentially learns a non-linear mapping \(\mathcal{F}\) between its inputs and outputs. The model is often called a ``network'' because it consists of billions of interconnected neurons, much like the biological neural networks in our brains.
\item
  Define a loss function \(\mathcal{L}\) that tells the network how bad the reconstruction is, and which way it should move to improve the reconstruction. Iterating through the dataset, the network nudges its billions of parameters to minimize this loss function. This is what we mean by ``learning'' in deep learning.
\item
  Test the network using data pairs not revealed during training. If the network performs well on unseen data, it can generalize to observational data with similar degradation.
\end{enumerate}

For translation-invariant data, the network in step. (iii) is often a Convolutional Neural Network (CNN) that learns to optimize layers of convolution kernels \citep{Lecun1989}. CNN is effectively a sparsely-connected neural network, where each neuron is only responsive to a restricted field of neurons in the previous layer. This field is often called the CNN's \emph{receptive field}. For a CNN with similar kernel sizes in each layer, the receptive field is small because the network is only sensitive to the scale of features that can be amplified by convolution kernels of a particular size. Efforts to enlarge the receptive field led to the development of the UNet architecture \citep{UNet2015}, which uses a \emph{contracting path} and an \emph{expansive path} to learn long-range spatial dependencies. The contracting path encodes the inputs by downsampling after each convolution, reducing spatial dimensions and iterating through different receptive field sizes. The expansive path then decodes the data by up-convolutions, while using skip connections to receive information from the contracting path. The U-shaped architecture is sensitive to different scales of features, and is thus very useful in image segmentation, classification, and translation tasks.

In the context of Dark Energy Survey (DES) Scientific Validation (SV), Jeffrey et al. designed the first UNet for weak lensing mass inversion (\href{https://github.com/NiallJeffrey/DeepMass}{DeepMass}; \citealt{Jeffrey2020ML}). The input shear data is first passed through the Wiener filter in Eq.~\((\ref{wf})\), then fed into UNet for non-linear refinement on the Wiener posterior to obtain the deep-learning reconstruction
\begin{equation}
    \bm{\hat{\kappa}}_\mathrm{UNet} = \mathcal{F}(\mathbf{W} \bm\gamma),
\end{equation}
where \(\mathcal{F}\) is learned by the network. The use of Wiener filtering compresses input shear to real space and reduces memory consumption, but in turn necessitates an input power spectrum in the form of a signal covariance matrix \(\mathbf{\Sigma_s}\). The loss function in step. (iv) is the mean square error (MSE) loss, or the squared \(l_2\) norm
\begin{equation}
    \mathcal{L}(\bm{\hat{\kappa}}_\mathrm{UNet}, \bm\kappa) = \mathrm{MSE} = 
    \mathbb{E} \left(
    \| \bm\kappa - \bm{\hat{\kappa}}_\mathrm{UNet} \|_2^2 \right), \label{mse1}
\end{equation}
which is the standard loss function in deep learning.

\section{MIU\textsuperscript{2}Net}\label{sec:MIU2Net}

We propose Mass-Inversion-U\textsuperscript{2}Net (MIU\textsuperscript{2}Net), a {\color{black}versatile} deep learning framework for weak lensing mass inversion. We adapt a nested UNet architecture (\href{https://github.com/xuebinqin/U-2-Net}{U\textsuperscript{2}-Net}; \citealt{U2Net2020}) for image translation tasks, and construct a new loss function to jointly optimize MSE and the frequency domain energy distribution. We introduce our simulation, network architecture and loss function in this section. Hyperparameters and technical details for training can be found in Appendix~\ref{sec:training-spec}.

\subsection{Training Data}\label{sec:Training Data}

Our training data is obtained from ray-tracing dark matter simulations. In order to cover the space between $0 < z \leq 1$, we concatenate $8$ pseudo-independent cubic boxes where each box has side length $320 \, \mathrm{h}^{-1} \mathrm{Mpc}$ and contains $640^3$ dark matter particles, each with mass $0.97 \times 10^{10} \, \mathrm{h}^{-1} \mathrm{M}_{\sun}$. The $8$ boxes share the same initial conditions, but are at different redshifts and are made pseudo-independent via translations and rotations. We have $4$ sets of boxes with different initial conditions (where each set contains $8$ boxes sharing the same initial conditions.) The cosmological constants used in the simulations are $\sigma_8 = 0.82$, $\Omega_\mathrm{m} = 0.28$, $\Omega_\mathrm{\Lambda} = 0.72$, $\Omega_\mathrm{b} = 0.046$, $h = 0.7$, and $n_s = 0.96$. 

In order to construct lensing systems, we divide each box into $5$ segments according to comoving distance along the line-of-sight. All particles in the segments are then projected onto 2D slices acting as lens planes. We have a total of $37$ lens planes between $0 < z \leq 1$ (the last box is not fully used). We then utilize multi-plane ray-tracing \citep{Hilbert2009raytrace} to obtain $6000$ pairs of shear maps $\gamma$ and convergence maps $\kappa$. The map sizes are $3.5 \times 3.5$ $\mathrm{deg}^2$ pixelated into $1024 \times 1024$ pixels.

We then prepare our training data for machine learning by dividing the data into $5000$ pairs of training data and $1000$ pairs of validation data. {\color{black} If we need to utilize reduced shear, we can easily compute a third set of reduced shear maps by equation (\ref{eq:reduced-shear}) from clean $\gamma - \kappa$ pairs. For fair comparisons, we omit this step and instead train our network on $\gamma - \kappa$ pairs} Then, we generate a shape noise realization corresponding to $20$ galaxies \(\rm{arcmin}^{-2}\) and add it to {\color{black}shear maps} according to equation (\ref{noise}). In principle, MIU\textsuperscript{2}Net sees a new training pair every time because the random shape noise realization completely dominates the shear field. After adding shape noise, we randomly crop each map to $512 \times 512$ pixels, before downsampling them to $256 \times 256$ to reduce memory consumption. To prevent over-fitting, we augment our data by random horizontal flips, vertical flips, and $90 \degr$ rotations. In this way, we obtain $320000$ independent pairs of training data and many more pairs with overlapped regions. {\color{black} To mimic bright star masks in observational settings, we generate a random bright star mask and apply it to the noisy reduced shear map every time MIU\textsuperscript{2}Net sees a training sample. The bright star masks consist of circular radii following a power law distribution and randomly cover $0 - 25\%$ of the pixel areas for every realization. By randomising mask percentage during training, the network can be versatile against different mask levels, and can reliably reconstruct convergence maps without retraining.}

% For better comparison with KS and WF, we first train a MIU\textsuperscript{2}Net with shear maps polluted only by shape noise. The results are outlined in Sect.~\ref{sec:Perceptual Quality}--\ref{sec:peak-count}.

% We then demonstrate MIU\textsuperscript{2}Net's ability to deal with reduced shear and data masks in the full training in Sect.~\ref{sec:full-miu2net-results}. In the full MIU\textsuperscript{2}Net training, we compute a third set of reduced shear maps by equation (\ref{eq:reduced-shear}) from clean $\gamma - \kappa$ pairs. We then add shape noise to the reduced shear maps, followed by cropping, downsampling, random flips and rotations. We also generate a random bright star mask and apply it to the noisy reduced shear map every time MIU\textsuperscript{2}Net sees a training sample. The bright star masks consist of circular radii following a power law distribution and cover $\sim 20 \%$ of the pixel areas for every realization. An example bright star mask is shown in the bottom left panel of Fig.~\ref{fig:shear}. 

\subsection{Architecture}\label{sec:Architecture}

MIU\textsuperscript{2}Net is based on U\textsuperscript{2}-Net originally designed for salient object detection \citep{U2Net2020}. U\textsuperscript{2}-Net is a nested UNet structure that greatly increases network depth without significantly increasing the computation cost. It has a large encoder-decoder U-structure, where each stage is a well-configured UNet with residual connections. Each decoder stage predicts a saliency probability map \(\mathcal{S}_\mathrm{side}^{(m)}\), and the final prediction is a weighted average of all \(\mathcal{S}_\mathrm{side}^{(m)}\) where the weights are learned via 1x1 convolution. In this way, U\textsuperscript{2}-Net can efficiently fuse information at all scales. The detailed design of U\textsuperscript{2}-Net is not the focus of this paper, and we encourage the reader to refer to the original paper and PyTorch implementation \href{https://github.com/xuebinqin/U-2-Net}{here}.

\begin{figure*}
    \centering
    \includegraphics[width=0.85\linewidth]{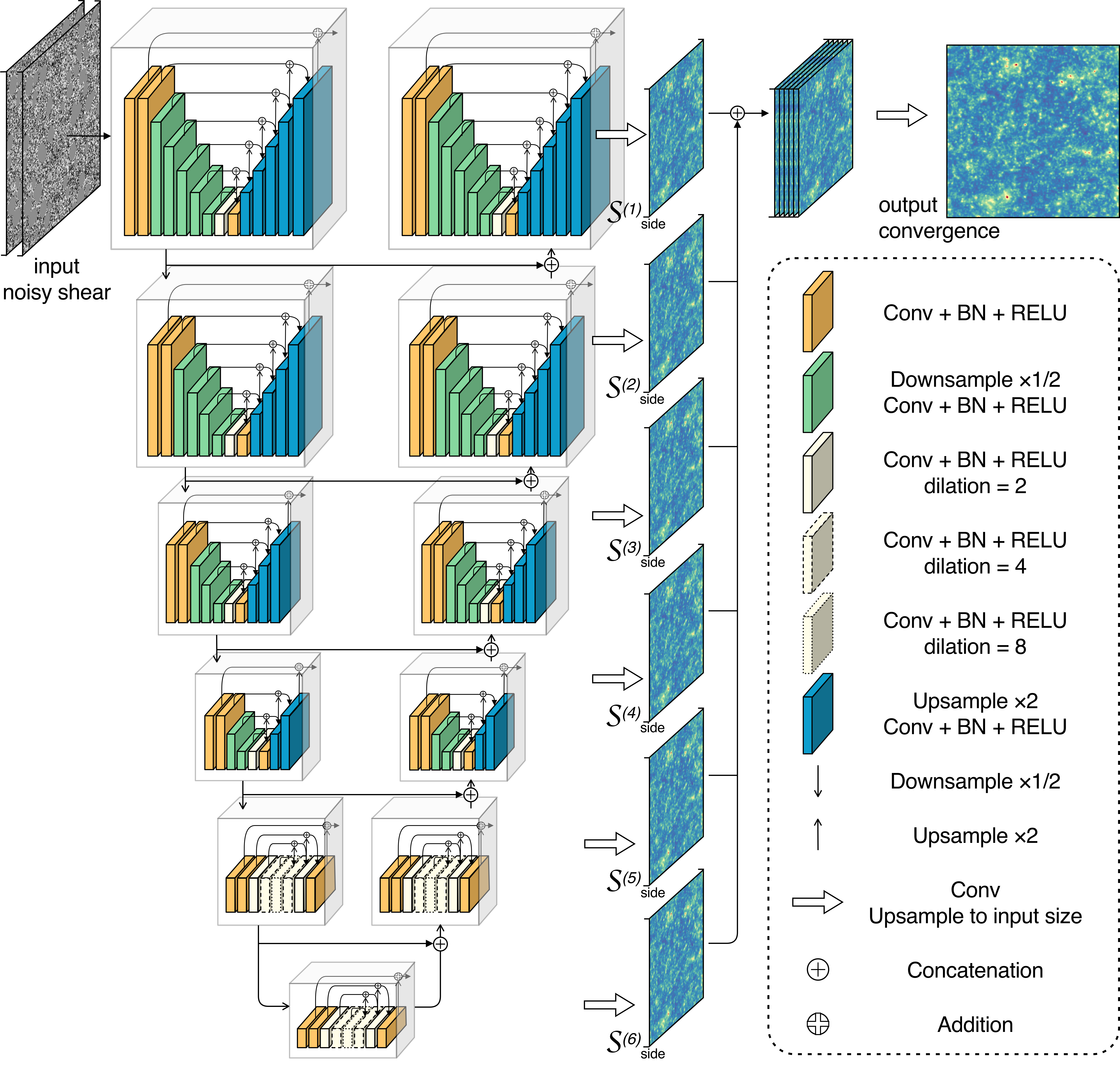}
    \caption{MIU\textsuperscript{2}Net structure adapted from U\textsuperscript{2}-Net. Figure is drawn based on Fig.~$5$ in \citet{U2Net2020}.}
    \label{fig:MIU2Net-network}
\end{figure*}

The original saliency probability maps fall between 0 and 1 because they denote the probability of each pixel belonging to the salient object. MIU\textsuperscript{2}Net removes the sigmoid function at the end of each decoder stage, so that the map intensities are no longer bounded and can correspond to true convergence maps. Fig.~\ref{fig:MIU2Net-network} shows the architecture of MIU\textsuperscript{2}Net in detail. During training, MIU\textsuperscript{2}Net takes the degraded complex reduced shear \(\bm g^{\mathbf n}\) as a two-channel input and outputs the true convergence map, learning a non-linear mapping
\begin{equation}
    \bm{\hat{\kappa}}_\mathrm{MIU^2Net} = \mathcal{F}(\bm g^{\mathbf n}).
\end{equation}

Further extensions to MIU\textsuperscript{2}Net are simple. We can fuse information from the projected halo distribution, for example, by simply appending the halo distribution map as a third channel to the input. The network will adjust its parameters to learn an optimal fusion.

\subsection{Loss Function}\label{sec:Loss Function}

The original U\textsuperscript{2}-Net has its loss function as a standard binary cross-entropy, because each pixel in the saliency probability maps is essentially a binary classification problem. To adapt it to our image translation task, we use a pixel-wise MSE loss as used by \citep{Jeffrey2020ML}. However, MSE-based networks are inclined to make conservative, blurry predictions averaged over all possible solutions weighted by their posterior probabilities, instead of learning the actual image manifold \citep{Blau2018PDtradeoff, Elad2023denoising}. This would make the minimum MSE (MMSE) result unrepresentative of the image distribution and influence global statistics. To obtain summary statistics from the MMSE solution, we encounter a similar problem as WF because the pixel values that form the optimized reconstruction $\bm{\hat\kappa}$ do not optimize its statistics $f(\bm{\hat\kappa})$. 

We approach this problem by adding a radial-averaged power spectrum (RAPS) term in the loss function \(\mathcal{L}\), so that we jointly optimize MSE and the frequency-domain reconstruction quality. The deep learning convergence map can then be used for inference and summary statistics. In addition, tuning the weights for different loss terms designates their relative importance and balances between optimal mean convergence and optimal summary statistic. It must be stressed, however, that a joint estimation is never the optimal estimation for each. One should tune their relative importance with respect to each application.

We denote the six decoder-stage maps as \(\bm{\hat\kappa}_\mathrm{side}^{(m)}\) and the final output map as \(\bm{\hat\kappa}_\mathrm{fuse}\). The loss function is computed between the true convergence map \(\bm\kappa\) and each of the decoder-stage maps, as well as between \(\bm\kappa\) and the final output map. During training, the loss function \(\mathcal{L}\) is the total loss of all seven maps, or
\begin{equation}
    \mathcal{L} = w_\mathrm{fuse} \, l_\mathrm{fuse} + 
    \sum_{m=1}^{6} w_\mathrm{side}^{(m)} \, l_\mathrm{side}^{(m)},
\end{equation}
where \(l_\mathrm{fuse}\) is the loss between \(\bm\kappa\) and \(\bm{\hat\kappa}_\mathrm{fuse}\), and \(l_\mathrm{side}^{(m)}\) is the loss between \(\bm\kappa\) and each of \(\bm{\hat\kappa}_\mathrm{side}^{(m)}\). \(w_\mathrm{fuse}\) and \(w_\mathrm{side}^{(m)}\) are weights for each loss term. We choose \(w_\mathrm{fuse}\) and all \(w_\mathrm{side}^{(m)}\) to be $1$. For each pair of true \(\bm\kappa\) and reconstructed \(\bm{\hat\kappa}\) where we compute a loss term \(l\), we have
\begin{equation}
    l = \alpha \, l_\mathrm{MSE} + \beta \, l_\mathrm{RAPS}, \label{loss}
\end{equation}
where \(l_\mathrm{MSE}\) and \(l_\mathrm{RAPS}\) are the MSE loss term and the RAPS loss term, respectively. \(\alpha\) and \(\beta\) determine their relative importance.

The MSE loss term \(l_\mathrm{MSE}\) is the standard MSE defined in Eq.~\((\ref{mse1})\). We rewrite it into a clearer form:
\begin{equation}
    l_\mathrm{MSE}(\bm\kappa, \bm{\hat\kappa}) = 
	\frac{1}{H \times W} \sum_{(i,j)}^{(H, W)}
	\left( \bm\kappa_{i,j} - \bm{\hat\kappa}_{i,j} \right)^2,
\end{equation}
where \((H, W)\) is the height and width of the convergence maps, here taken to be \((256, 256)\).

The RAPS loss term \(l_\mathrm{RAPS}\) quantifies reconstruction quality in the frequency domain. For each pair of \(\bm\kappa\) and \(\bm{\hat\kappa}\), we take their Discrete Fourier Transforms and compute their respective power spectra \(P_{\bm\kappa}(u, v)\) and \(P_{\bm{\hat\kappa}}(u, v)\). We define the radial-average power spectrum as
\begin{equation}
    \bar{P}_{\bm\kappa}(r) = \frac{1}{N(r)}
    \sum_{r} P_{\bm\kappa}(u, v),
\end{equation}
where the summation is performed over all Fourier-space coordinates at a distance \(r\) from the DC component, i.e., over all \((u, v)\) pairs that satisfy \(r-1 \leq \sqrt{u^2 + v^2} < r\). \(N(r)\) denotes the number of \((u, v)\) pairs within this annulus. In practice, it is impossible to reconstruct the high-frequency fluctuations under realistic noise, so we limit RAPS to low frequencies (central regions in Fourier space) and only consider \(r\) values up to \(r_{\max}\). We choose \(r_{\max} = 16.0\) in this paper. The RAPS loss is then defined as the Mean Absolute Error (MAE) between central regions of true and predicted radial-average power spectra, or
\begin{equation}
    l_\mathrm{RAPS}(\bm\kappa, \bm{\hat\kappa}) =
	\frac{1}{r_\mathrm{max}} \sum_{r=0}^{r_\mathrm{max}}
	\left\lvert \bar{P}_{\bm\kappa}(r) - \bar{P}_{\bm{\hat\kappa}}(r)
	\right\lvert. \label{raps}
\end{equation}

% As detailed in Sect.~\ref{sec:Deep Learning}, the MSE term is a strong estimator for the mean convergence, and the RAPS term places a tight constraint on the relevant summary statistics. In this way, the combined loss function guides the network to break the mass-sheet degeneracy and reconstruct the convergence power spectrum simultaneously.
%\ranli{We may need to discuss the parts regarding mass-sheet.}

\section{Results}\label{sec:Results}

{\color{black}In this section, we compare our convergence maps to those obtained by established methods, including unsupervised methods such as Kaiser-Squires (KS), Wiener filtering (WF), MCALens (MCA) \citep{Starck2021MCALens}, and supervised deep learning methods such as DeepMass (UNet) \citep{Jeffrey2020ML}. For easier comparison with established methods, we first test MIU\textsuperscript{2}Net with shear maps polluted only by shape noise, i.e., using the degraded shear field \(\bm\gamma^{\mathbf n}\) instead of the reduced shear \(\bm g^{\mathbf n}\), and adding no data masks. The results are outlined in Sect.~\ref{sec:Perceptual Quality}--\ref{sec:rmse}. We then demonstrate MIU\textsuperscript{2}Net's {\color{black}generalization across cosmologies in Sect.~\ref{sec:cosmo2}, and its} ability to deal with data masks in the full training in Sect.~\ref{sec:full-miu2net-results}. In all test cases, we use the same trained model without further retraining or fine-tuning. 

For DeepMass reconstructions, because the original implementation is designed for DES SV mass inversion, we replicate the authors' network structure and retrain the model on our simulation for 512 epochs (network converges at about 200 epochs). {\color{black} Because our replication is not the same as the official DeepMass, we change the name to ``SimpleUNet'' or the shorthand ``UNet''.} For MCALens, we use $\lambda = 5$ and \textbf{InpWiener} for inpainting, as illustrated in \citet{Starck2021MCALens}. For both Wiener filtering and MCAlens, we input the correct signal power spectrum into the algorithms. There are also many variants of these established methods (e.g. inpainting using different priors, as in \citet{Pires2009KS+, Lanusse2016Glimpse, Pires2020KS+, Starck2021MCALens}, and iterative E-mode convergence estimation, as in \citet{SchneiderSeitz1995-II, Jullo2014}).} For future studies, a thorough comparison is desired to characterize MIU\textsuperscript{2}Net relative to each of the traditional variants.

\subsection{Perceptual Quality}\label{sec:Perceptual Quality}
% \ranli{To Wenhan: In this part, we focus on the visual assessment, and we might want not to be too absolute in our statements to avoid prompting reviewers to demand quantitative validation. We can speak in general terms about the perceived differences, as we have enough figures later to substantiate the benefits of our approach.}

{\color{black}Fig.~\ref{fig:visual} shows a visual comparison between MIU\textsuperscript{2}Net, KS, WF, MCA, and {\color{black}UNet} reconstructions.} A few things are immediately noticeable:

\begin{enumerate}
\item
  MIU\textsuperscript{2}Net has the best perceptual quality and most accurately represents the data distribution. It has superior reconstruction quality from large-scale, filament-like structures to well-defined, non-Gaussian peaks.
\item
  Compared to KS reconstruction, MIU\textsuperscript{2}Net is highly effective in suppressing noise. 
\item
  MIU\textsuperscript{2}Net reconstruction is very similar to WF in terms of large-scale diffusive structures. Although we do not feed WF into MIU\textsuperscript{2}Net in any way, MIU\textsuperscript{2}Net seems to have learned the WF result.
\item
  MIU\textsuperscript{2}Net reconstructions have well-defined local peaks. The peak centroids are close to the true centroids and can potentially be used to locate hidden dark matter halos.
\end{enumerate}

\begin{figure*}[!h]
    \centering
    \includegraphics[width=0.75\linewidth]{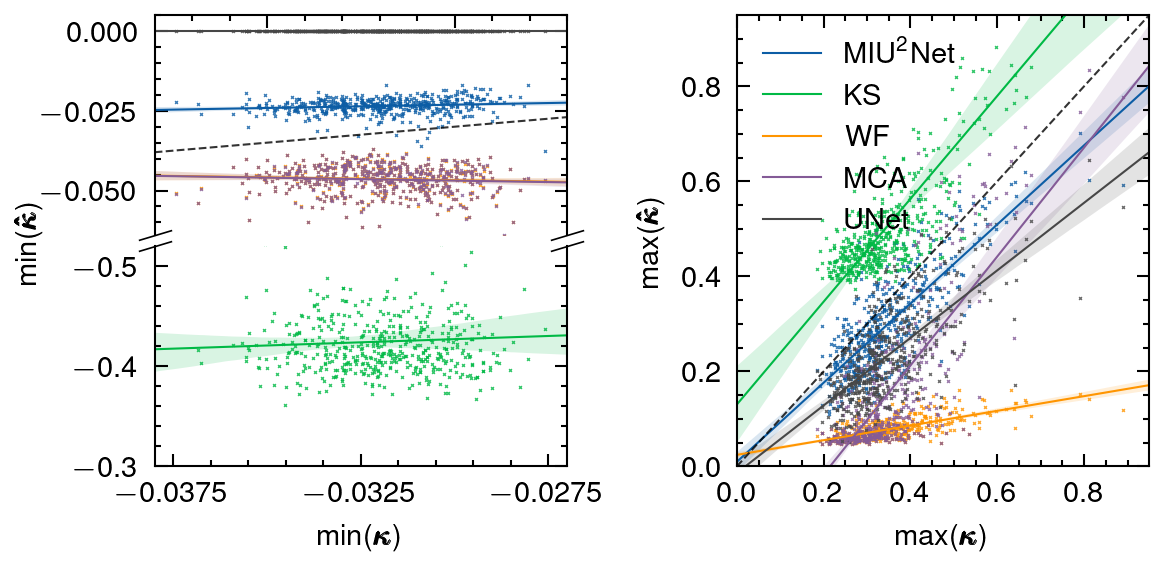}
    \caption{{\color{black}Dynamic range for each of the \(500\) convergence reconstructions from MIU\textsuperscript{2}Net, KS, WF, MCA, and {\color{black} UNet}.} \emph{Left panel}: minimum predicted convergence value \(\min(\bm{\hat\kappa})\) against minimum true convergence value \(\min(\bm\kappa)\) for each reconstruction. {\color{black} We also plot the best-fit lines, and the shadings indicate the \(95\%\) confidence intervals for each method.} The coloured dots follow the same labelling scheme as in the right panel. The black dashed line is the \(y=x\) line denoting ideal reconstruction. Note that MIU\textsuperscript{2}Net can best recover the lower bound, but it systematically overestimates it when \(\min(\bm\kappa)\) is low. KS reconstructions cluster around the lower region far away from truth, because KS is symmetric around \(0\) and does not provide a good estimate for the lower bound. \emph{Right panel}: {\color{black} scatter plot and best-fit lines for maximum values for each true-prediction pair. The shadings indicate the \(95\%\) confidence intervals for each method.} Again, the black dashed line is the \(y=x\) line tracing the ideal reconstruction. It is clear that MIU\textsuperscript{2}Net has a slope closest to the ideal and can reliably predict the upper bound of individual convergence maps. Nevertheless, MIU\textsuperscript{2}Net systematically underestimates the upper bound. This is expected because an MSE-based estimator is inclined to make blurry, middle-range predictions that smooth out intense spikes. {\color{black}Compared to {\color{black} UNet}, which is also an MSE-based estimator, MIU\textsuperscript{2}Net more reliably recovers the maximum amplitude. As seen by the larger slope, MIU\textsuperscript{2}Net is more inclined to make bold, salient predictions than {\color{black} UNet} for larger maximum peaks.}}
    \label{fig:dynamic range}
\end{figure*}

Part of the good perceptual quality arises from MIU\textsuperscript{2}Net's ability to recover the true dynamic range of convergence maps. {\color{black}Fig.~\ref{fig:dynamic range} shows the dynamic range recovery for MIU\textsuperscript{2}Net, KS, WF, MCA, and {\color{black}UNet}.} MIU\textsuperscript{2}Net significantly outperforms other methods in recovering the lower and upper bounds of individual convergence maps. Importantly, the upper bound \(\max(\bm\kappa)\) of a given convergence map corresponds to the pixel with the most intensity, and therefore with the most concentrated projected matter density in a region. Accurate recovery of the position and amplitude of \(\max(\bm\kappa)\) is crucial for finding hidden mass concentrations that are absent in the electromagnetic space.

%Because weak lensing mass mapping is sensitive to all matter, it is a powerful probe for dark matter halos that undergo little baryonic interactions. Cross-correlations between weak lensing surveys and optical, IR, X-ray methods may locate baryon-deficient halos and reveal important properties of dark matter.

\begin{figure*}
    \centering
    \includegraphics[width=0.7\linewidth]{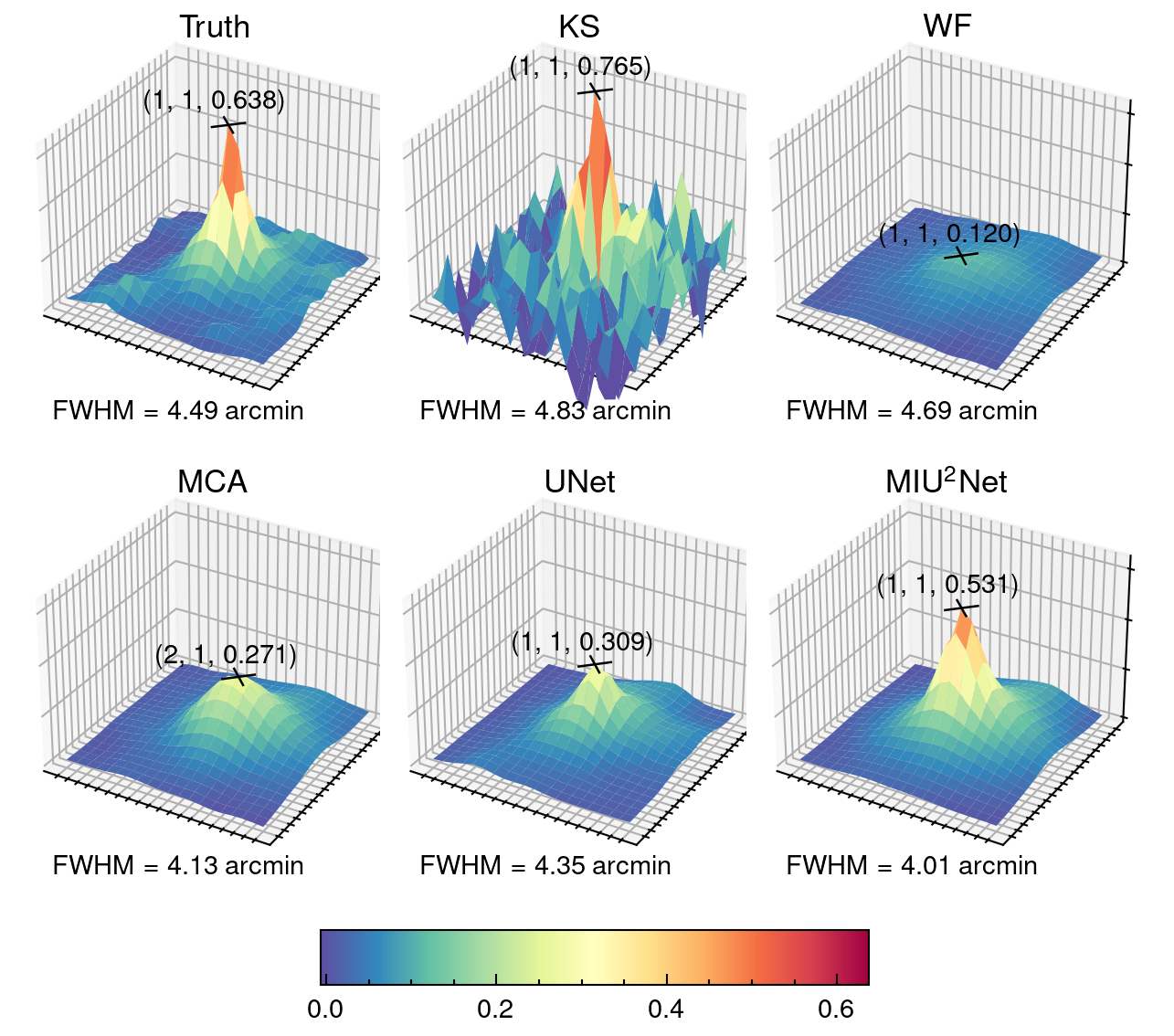}
    \caption{{\color{black}Centroid and amplitude recovery from MIU\textsuperscript{2}Net, KS, WF, MCA, and {\color{black} UNet} for an individual convergence peak.} The plotted regions have dimensions \(8.2 \times 8.2 \, \rm arcmin^2\) and correspond to the scarlet dashed square regions in Fig.~\ref{fig:visual}. The crosses indicate the highest peak in each of the recovered regions. The coordinates above the crosses are in the form \((\Delta x, \Delta y, \kappa_{\rm peak})\), where \(\Delta x\) and \(\Delta y\) are spatial coordinates of the predicted peak relative to that of the true peak, and \(\kappa_{\rm peak}\) is the peak convergence value. The Full Width at Half Maximum (FWHM) for each peak is shown in the bottom left of each panel. MIU\textsuperscript{2}Net best recovers the position, amplitude, and FWHM of the true peak {\color{black}without significant noise}. However, MIU\textsuperscript{2}Net fails to recover the highly non-Gaussian profile of the true peak, as the reconstruction still exhibits a relatively smoothed profile {\color{black} and a bell-like shape. Note that although KS recovers the central peak with decent accuracy, this is because the central peak has a large convergence value ($0.638$). For lower convergence values, noise can easily submerge the reconstructed peak.}}
    \label{fig:profile3d}
\end{figure*}

We also visualize an example mass density peak to show MIU\textsuperscript{2}Net's reconstruction of peak centroid and amplitude. {\color{black}Fig.~\ref{fig:profile3d} shows a mass density peak in the true convergence map, as well as the same regions in the reconstructions by MIU\textsuperscript{2}Net, KS, WF, MCA, and {\color{black} UNet}. Although KS has a decent reconstruction of the central intensity, it is very noisy and less sensitive for lower convergence values.} Compared to {\color{black}WF, MCA, and {\color{black} UNet}}, the MIU\textsuperscript{2}Net peak is much more prominent; its amplitude is also much closer to the true amplitude. MIU\textsuperscript{2}Net also has optimal peak recovery based on Full Width at Half Maximum (FWHM) measures. However, the MIU\textsuperscript{2}Net peak profile is still relatively smooth, and it fails to reproduce the highly non-Gaussian profile of a true convergence peak.

\subsection{On the Mass-Sheet
Degeneracy}\label{sec:breaking-the-mass-sheet-degeneracy}

\begin{figure*}
    \centering
    \includegraphics[width=1\linewidth]{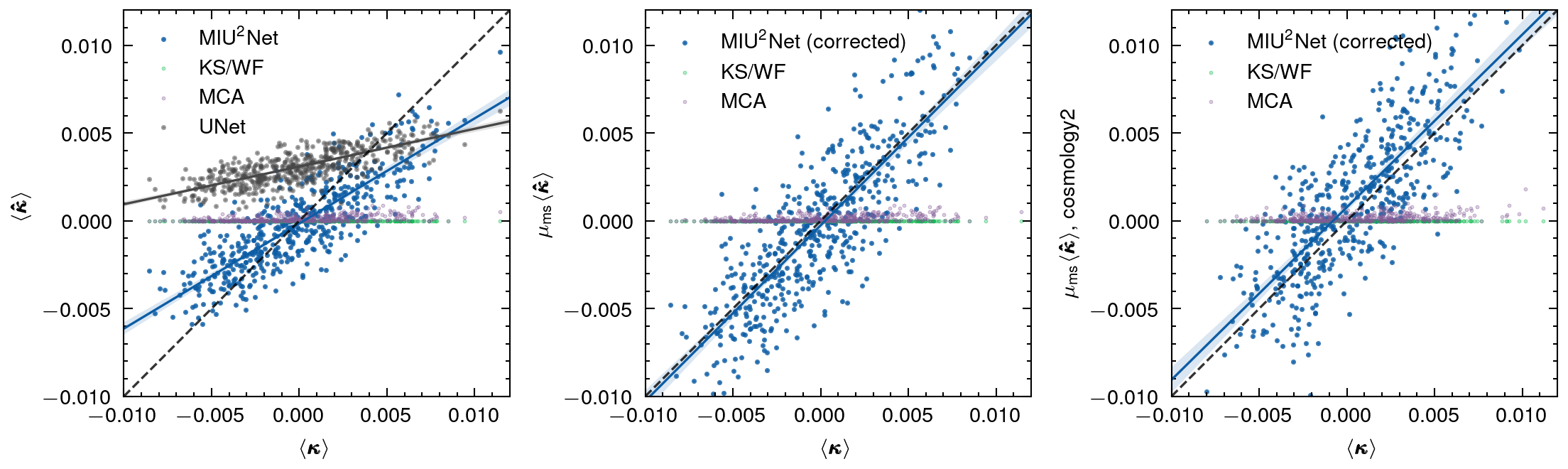}
    \caption{On the mass-sheet degeneracy. \emph{Left panel}: for \(500\) reconstructions, we plot the mean \(\langle \bm{\hat\kappa} \rangle\) for each of the predicted convergence maps against the mean \(\langle \bm\kappa \rangle\) for each of the true convergence maps. The black dashed line is the \(y = x\) line denoting the ideal recovery of mean convergence. MIU\textsuperscript{2}Net reconstructions clearly follow a linear relation whose slope is close to the ideal and has a negligible offset. The blue line and shading indicate the best-fit line for MIU\textsuperscript{2}Net and the \(95\%\) confidence interval. Both KS and WF assume \(\langle \bm{\hat\kappa} \rangle = 0\), so they overlap on the \(y = 0\) line. {\color{black} {\color{black} UNet} also exhibits a positive slope, suggesting that deep learning MMSE estimators may utilize the prior information in convergence fields better than other methods.} \emph{Middle panel}: the same as left panel, except now we plot the corrected MIU\textsuperscript{2}Net reconstructions with \(\mu_{\rm ms} = 1.672\). The value of \(\mu_{\rm ms}\) is chosen empirically based on the parameters of the best-fit line in the left panel. The best-fit line for corrected MIU\textsuperscript{2}Net closely resembles the ideal linear relationship. {\color{black}\emph{Right panel}: the corrected MIU\textsuperscript{2}Net reconstructions with the same scaling factor \(\mu_{\rm ms} = 1.672\) for reconstructions using the second cosmology. The corrected mean convergence is still very close to the ideal reconstruction despite the change in cosmology.}}
    \label{fig:mass sheet}
\end{figure*}

As shown in Eq.~\((\ref{eq:mass-sheet})\), mass-sheet degeneracy prohibits us from recovering the mean convergence from the ellipticity field. MIU\textsuperscript{2}Net, however, appears to be capable of inferring the mean convergence using only the shear field. This may be primarily due to the prior of the training set used. As shown in Fig.~\ref{fig:mass sheet}, instead of assuming mean convergence to vanish, MIU\textsuperscript{2}Net recovers the mean with some scatter. The slope of the best-fit line is less than \(1\), which is expected because MSE-based loss functions are inclined to produce conservative, middle-ranged reconstructions. {\color{black} Both MIU\textsuperscript{2}Net and {\color{black} UNet} exhibit a positive slope, but MIU\textsuperscript{2}Net recovers the mean convergence better and has a negligible overall bias. We thus introduce the simple correction factor $\mu_{\rm ms} = 1.672$, so that the corrected MIU\textsuperscript{2}Net statistics \(\mu_{\rm ms} \langle \bm{\hat\kappa} \rangle\) matches the ideal recovery, as shown in the middle panel of Fig.~\ref{fig:mass sheet}. Interestingly, although we only introduce a scaling factor, it is robust under a significantly different cosmology (right panel in Fig.~\ref{fig:mass sheet}) and $20\%$ masked conditions (see Fig.~\ref{fig:mg_mass sheet}). The second cosmology is outlined in detail in Sect.~\ref{sec:cosmo2}. This suggests that our empirical correction factor obtained from a single, fiducial cosmology generalizes across cosmologies for reconstructing dark matter environment.} Many methods have been proposed to lift the mass-sheet degeneracy \citep{SeitzSchneider1997-III, BradacSchneider2004MassSheet, Rexroth2016Moments, Cremonese2021GW}; with additional data such as redshift distribution, flexion fields, and gravitational wave inferences, MIU\textsuperscript{2}Net {\color{black}and deep learning estimators} offer the prospect of breaking the mass-sheet degeneracy and probing the absolute mass density distribution. 

% This may allow us to probe the absolute mass density distribution from weak lensing alone. Because the mean of each MIU\textsuperscript{2}Net reconstructed map is non-zero, MIU\textsuperscript{2}Net mass mapping may be uniquely capable of detecting massive-scale dark matter inhomogeneities on sky. Peaks recovered in the convergence field also have absolute amplitudes that directly translate to the mass densities of prominent dark matter haloes along the lines of sight weighted by the lensing efficiency kernel.

\subsection{Convergence Distribution}\label{sec:convergence-distribution}

\begin{figure*}
    \centering
    \includegraphics[width=1\linewidth]{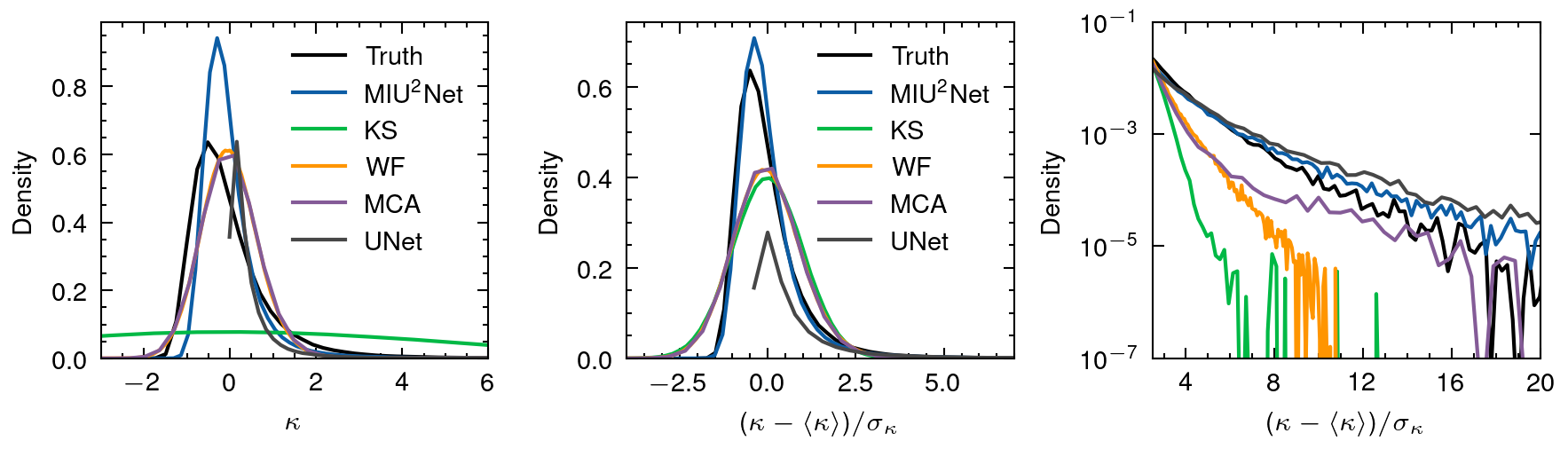}
    \caption{{\color{black}Convergence probability density distribution for MIU\textsuperscript{2}Net, KS, WF, MCA, and {\color{black} UNet}.} \emph{Left panel}: Kernel Density Estimation (KDE) for \(500\) reconstructions. No method appears to recover the true distribution, but the MIU\textsuperscript{2}Net distribution follows a log-normal curve similar to the truth. \emph{Middle panel}: KDE for the same \(500\) reconstructions normalized by the mean and standard deviation of all constituent pixels. The normalized MIU\textsuperscript{2}Net closely resembles the true distribution, whereas {\color{black}KS, WF, and MCA} are symmetric around \(0\) and do not exhibit any log-normal shape. {\color{black}{\color{black} UNet} has a significant amount of pixels around $\kappa = 0$ and does not follow the true distribution.} \emph{Right panel}: a zoom-in for the extended high-end tail for normalized distributions. The \(y\)-axis is in log scale for better visualization. Note that KS and WF fall off rapidly, whereas MIU\textsuperscript{2}Net recovers the extended tail with slight overestimation at the high end.}
    \label{fig:distribution}
\end{figure*}

The distribution of convergence values follows a log-normal distribution with an extended tail \citep{Jain2000tail,  Hilbert2011tail, Clerkin2017tail}. Fig.~\ref{fig:distribution} shows the convergence distribution in terms of Kernel Density Estimation (KDE) for truth and all reconstruction methods. In the left panel, we show that none of the methods recovers the true distribution well, but MIU\textsuperscript{2}Net has a log-normal shape similar to that of the truth. We then transform each reconstruction method by the normalization \(\kappa \rightarrow (\kappa - \langle \kappa \rangle) / \sigma_\kappa\), where \(\langle \kappa \rangle\) and \(\sigma_\kappa\) are the mean and standard deviation of all constituent pixels in all maps reconstructed by that method. The distribution for normalized convergence values for each method is shown in Fig.~\ref{fig:distribution}, middle panel. The normalized MIU\textsuperscript{2}Net properly recovers the true distribution, whereas {\color{black}KS, WF, and MCA} are symmetric and cannot recover the log-normal curve. {\color{black}{\color{black} UNet} has a large portion of pixels around $\kappa 
= 0$ and does not follow the true distribution as well.} The right panel zooms in on the high-end extended tail largely resembling convergence peaks. We show that KS and WF fall off rapidly, yet MIU\textsuperscript{2}Net closely follows the extended tail to very high convergence values until the true distribution itself starts to exhibit high variance. MIU\textsuperscript{2}Net's slight overestimation for the extended tail can be attributed to its imperfect recovery of non-Gaussian peak profiles, as seen qualitatively in Fig.~\ref{fig:profile3d}. Because a MIU\textsuperscript{2}Net peak does not rise as steeply as a true peak, extra intensity around the peak centroid may lead to overestimated probability density at middle and high convergence values. 

\subsection{Power Spectrum}\label{sec:power-spectrum}

One of the most important aspects of weak lensing studies to this date has been the measurement of two-point statistics \citep{Hildebrandt2017twopt, Troxel2018twopt, Hikage2019twopt, Hamana2020twopt}. For a zero-mean Gaussian density field, two-point statistics like the power spectrum or two-point correlation functions can completely characterize the data. However, systematic studies on popular methods such as KS, WF, null B-mode prior, and sparsity prior (\href{https://github.com/CosmoStat/Glimpse/tree/v1.0}{Glimpse}; \citealt{Lanusse2016Glimpse}) have shown that none of these methods can reliably reconstruct the power spectrum \citep{Jeffrey2021AllMethods}. Importantly, these methods differ in their priors used, but all are MAP estimators for the most probable map \(\bm{\hat\kappa}\) instead of the most probable summary statistic. Current deep learning methods, on the other hand, often focus on pixel-wise, MMSE-only reconstruction that does not translate well to the frequency domain. It has also been shown that a conditional generative adversarial network incorporating a learned loss function still cannot accurately recover the power spectrum under a more lenient noise per pixel than what is used in this paper \citep{Shirasaki2019ML}.

\begin{figure*}
    \centering
    \includegraphics[width=0.8\linewidth]{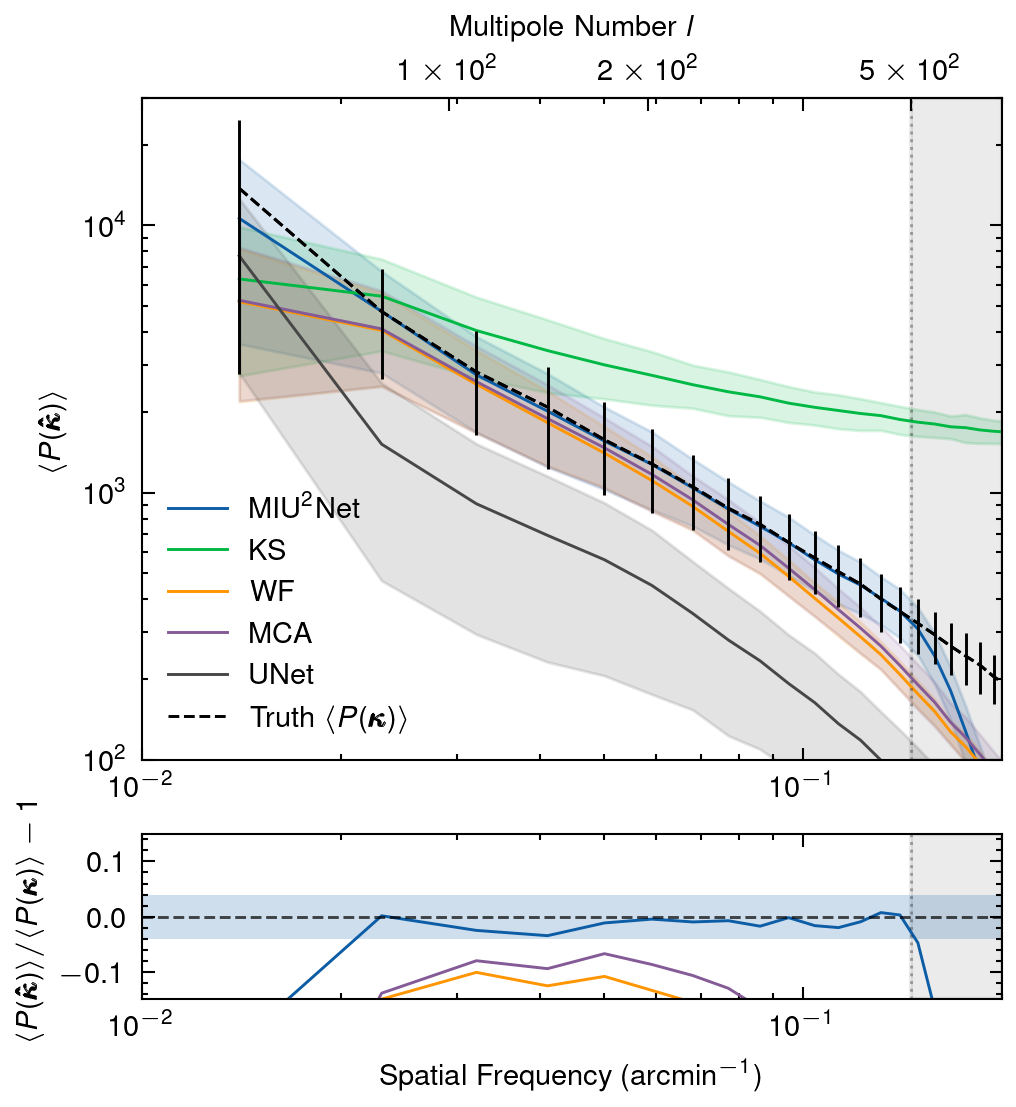}
    \caption{Power spectrum comparison. \emph{Top panel}: {\color{black}radial-averaged power spectra for truth, MIU\textsuperscript{2}Net, KS, WF, MCA, and {\color{black} UNet}}. The black dashed line and error bars show the mean and the standard deviation of the true power spectrum for \(500\) realizations. For each of the reconstruction methods, the coloured lines show the mean power spectrum \(\langle P(\pmb{\hat\kappa}) \rangle\) and the shadings mark the respective standard deviation. The grey vertical dotted line denotes the cutoff \(r_{\max}\) for MIU\textsuperscript{2}Net's RAPS loss, and the grey region to the right corresponds to high frequencies \emph{not} included in our implementation of RAPS loss. MIU\textsuperscript{2}Net power spectrum closely follows the true mean and standard deviation within the range of RAPS loss. The discrepancies between the true power spectrum and all recovered power spectra at the lowest frequency bin can be partly attributed to the imperfect recovery of the DC component, i.e. mean convergence. At high frequencies going into the grey region, the power spectrum for MIU\textsuperscript{2}Net rapidly diverges from the truth, as those frequencies are not considered in our loss function. Choosing different cutoff radii \(r_{\max}\) in the frequency domain changes the extent of the grey region. \emph{Bottom panel}: residual power spectra for each reconstruction method with respect to the truth. The coloured lines correspond to the same label in the top panel, but {\color{black}KS, and {\color{black} UNet}} are now out of the plot range. The blue band bounds the region \([-0.04, 0.04]\) where most of the MIU\textsuperscript{2}Net residual power spectrum resides. Note that MIU\textsuperscript{2}Net slightly underestimates the power at almost all frequencies, but it always stays within the \(\pm 4\%\) uncertainty band except at the lowest frequency bin.}
    \label{fig:power spectrum}
\end{figure*}

In Sect.~\ref{sec:Loss Function}, we have designed a combined loss function that enables MIU\textsuperscript{2}Net to jointly estimate a pixel-wise reconstruction and the frequency domain power distribution. We demonstrate in Fig.~\ref{fig:power spectrum} that MIU\textsuperscript{2}Net can reconstruct the convergence power spectrum with \(4\%\) uncertainties in the multipole range \(0 < l \lesssim 500\), far exceeding existing reconstruction methods at all scales. Compared to WF {\color{black}and MCA} with optimal input signal power spectra, MIU\textsuperscript{2}Net achieves \(1060\%\) {\color{black}and $930\%$} improvements in reconstruction quality at \(l \simeq 500\). Changing the high-frequency cutoff \(r_{\max}\) in Eq.~\((\ref{raps})\) changes the highest multipole number \(l\) considered in the loss function, so MIU\textsuperscript{2}Net can potentially reach even smaller angular scales with careful tuning. At the lowest frequency bin where recovering the DC component is subjected to the mass-sheet degeneracy, MIU\textsuperscript{2}Net has \(-20.2\%\) reconstruction error, whereas KS and WF have \(-54.4\%\) and \(-62.5\%\) error in the same frequency bin, respectively. 

The effect of intrinsic alignment and other systematic becomes important when we can constrain the power spectrum to within \(\pm 4\%\) up to \(l \simeq 500\). Intrinsic alignment can introduce several percent of uncertainties in two-point statistics \citep{Joachimi2013IA, Pires2020KS+}, but we do not yet have simulations that can model and study intrinsic alignment in detail. Other sources of error such as PSF modelling, photometric redshift, shear estimation, baryonic effects, and data preprocessing should also receive more attention as they start to become more prominent in the error budget. {\color{black}Although the error in power spectrum reconstruction ($4\%$) is yet insufficient for constraining cosmological parameters, regularising our MMSE estimator in the frequency domain can encourage the network to predict the right amount of peaks of different scales, especially peaks with higher spatial frequencies. Compared to pure MMSE estimators that tend to produce blurry, averaged predictions, MIU\textsuperscript{2}Net can better match the amplitude of large convergence peaks, as well as predict areas of intermediate convergence values from faint shear signals. This can greatly help map the distribution of large-scale structures in the universe, where filaments are usually fainter than local overdensities. }{\color{black} Compared to non-MMSE deep learning networks, such as conditional GAN variants with a trained loss function \citep{Shirasaki2019ML}, frequency-domain regularization may offer more advantage in estimating the convergence power spectrum. Our method can thus open many possibilities, including galaxy evolution studies under different dark matter environments (i.e. filaments, voids, peaks). In the future, improved training datasets with multiple cosmologies can greatly enhance our reconstructions and enable MIU\textsuperscript{2}Net to constrain cosmological parameters.}

% =======================================================
{\color{black}

\subsection{RMSE}\label{sec:rmse}

We also evaluate our reconstruction results with Root Mean Square Error (RMSE) under different smoothing conditions. {\color{black}Let
\begin{itemize}
    \item $\text{Ind} = \{ i \mid M[i] = 1 \}$ be the indices of the mask,
    \item $M$ be the binary mask (with the same shape as image),
    \item $G_\sigma(\cdot)$ be the Gaussian smoothing operator with a smoothing kernel of FWHM $\sigma$,
    \item $\bm\kappa, \bm{\hat{\kappa}}$ be the truth and reconstructed convergence maps, respectively.
\end{itemize}
Then Define
\begin{align*}
    \quad & Z_\sigma = M \cdot G_\sigma(M \cdot \bm\kappa), \\
    \quad & Z_\sigma[\text{Ind}] = Z_\sigma[\text{Ind}] - \text{mean}(Z_\sigma[\text{Ind}]), \\
    \quad & X_\sigma = M \cdot G_\sigma(M \cdot \bm{\hat{\kappa}}), \\
    \quad & X_\sigma[\text{Ind}] = X_\sigma[\text{Ind}] - \text{mean}(X_\sigma[\text{Ind}]),
\end{align*}
So we have the normalized RMSE
\begin{equation}
\text{RMSE}(\sigma) = \sqrt{ \frac{ \sum_{\text{Ind}} (Z_\sigma - X_\sigma)^2 }{ \sum_{\text{Ind}} Z_\sigma^2 } }.
\end{equation}
In this way, we can calculate RMSE without taking into account the mean value of the convergence field and the masked areas.} Note that by standard practices KS requires a smoothing on small angular scales, which is itself also a free parameter. Fig.~\ref{fig:rmse} shows that MIU\textsuperscript{2}Net achieves significantly lower RMSE than established methods across all smoothing scales. At larger smoothing scales, MIU\textsuperscript{2}Net is increasingly more accurate than other methods, demonstrating that MIU\textsuperscript{2}Net has a precise reconstruction of the underlying low-frequency component of the convergence field.
% \begin{equation}
%     \mathrm{RMSE}(\sigma) \equiv
%     \sqrt{\sum_{i=1}^N \left( 
%     G_\sigma ( \bm\kappa_i - \bm{\hat\kappa}_i
%     ) \right)^2
%     / \sum_{i=1}^N \bm\kappa_i^2},
% \end{equation}
% where $\bm{\kappa}$ is the true convergence map, $\bm{\hat\kappa}$ is the reconstruction, $N = 500$ is the number of reconstructed convergence maps, and $G_\sigma$ is the Gaussian smoothing function with a smoothing kernel of FWHM $\sigma$. 

At each smoothing level, we quantify our improvement by
\begin{equation}
    \mathrm{Imp} = \left( \mathrm{RMSE}_i - \mathrm{RMSE_{MIU^2Net}} \right)
    / \mathrm{RMSE}_i,
\end{equation}
where $\mathrm{RMSE_{i}}$ is the RMSE of a reconstruction method. Without smoothing, MIU\textsuperscript{2}Net is $\mathrm{Imp} = 83\%$ better than KS, and $\mathrm{Imp} = 5\%$ better than {\color{black} UNet}. With $1$ arcmin smoothing applied, KS is significantly improved, and MIU\textsuperscript{2}Net offers $\mathrm{Imp} = 34\%$ improvement over KS, and $\mathrm{Imp} = 38\%$ improvement over {\color{black} UNet}. MIU\textsuperscript{2}Net achieves similar levels of improvement to WF, and MCA. MIU\textsuperscript{2}Net also consistently outperforms {\color{black} UNet} on RMSE despite both being MMSE estimators. Changing the network structure and regularising in the frequency domain are therefore effective techniques for mass inversion.

\begin{figure*}
    \centering
    \includegraphics[width=0.6\linewidth]{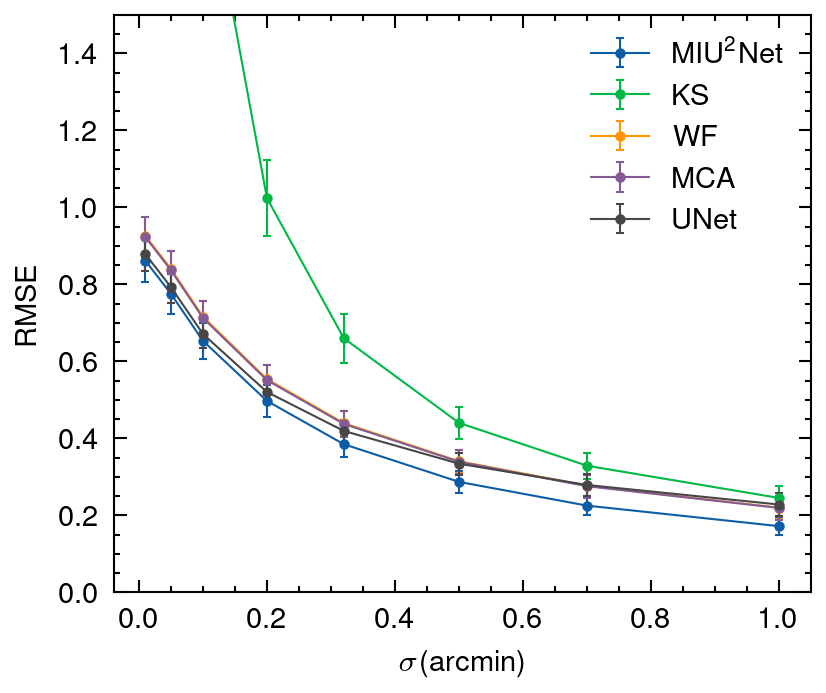}
    \caption{{\color{black}Root Mean Square Error (RMSE) under different smoothing conditions for each reconstruction method. Error bars denote one standard deviations. It is clear that MIU\textsuperscript{2}Net significantly outperforms all other methods across all smoothing scales.}}
    \label{fig:rmse}
\end{figure*}

\subsection{Generalization Across Cosmologies}\label{sec:cosmo2}

In this section, we show MIU\textsuperscript{2}Net's ability to generalize across cosmologies despite being trained on a single cosmology. As outlined in Sect.~\ref{sec:Training Data}, we trained our network with $\sigma_8 = 0.82$, $\Omega_\mathrm{m} = 0.28$, $\Omega_\mathrm{\Lambda} = 0.72$, $\Omega_\mathrm{b} = 0.046$, $h = 0.70$, and $n_s = 0.96$. We generate another 500 shear-convergence pairs with our simulation using $\sigma_8 = 0.81$, $\Omega_\mathrm{m} = 0.31$, $\Omega_\mathrm{b} = 0.049$, $\Omega_\mathrm{c} = 0.26$, $\Omega_\mathrm{\nu} = 0.0014$, $h = 0.68$, and $n_s = 0.97$ (fiducial in \cite{ChenZhao25}). We use our trained MIU\textsuperscript{2}Net model to directly predict convergence maps without fine-tuning. 

Fig.~\ref{fig:cosmo2 power spectrum} shows the power spectrum reconstruction for the second cosmology. In the top panel, we show the radial-averaged power spectrum for the new cosmology in black dashed line, and the old one in black dotted line. Note that the graph is a log-log plot, and the power spectrum difference between the two cosmologies is significant, reaching $19\%$ at the greatest. The lower panel shows the residual power spectrum for the original cosmology in black dotted line. Although we trained MIU\textsuperscript{2}Net on the original cosmology and regularised it with the dotted power spectrum, MIU\textsuperscript{2}Net is able to move away from the trained power spectrum and progress significantly closer to the new truth (black dashed line in both panels). Note that the power spectrum under the new cosmology has higher energy across all frequencies, and it is uncommon for an MMSE estimator to risk a prediction with high energy, especially when regularised on a power spectrum with lower energy. This demonstrates that the MSE term in our loss function is powerful enough to bring up the power spectrum when needed, and our joint loss function is beneficial for both MSE minimisation and power spectrum estimation. Compared to WF and MCA with the correct cosmology as inputs, MIU\textsuperscript{2}Net reaches the same accuracy in intermediate spatial frequencies and better accuracies in low and high spatial frequencies. Quantitatively, MIU\textsuperscript{2}Net achieves $10.8\%$ power spectrum accuracy across all frequency ranges when regularised on a different cosmology, $390\%$ better than Wiener filtering and $350\%$ better than MCALens at $l \simeq 500$. In real observational scenarios, we can retrain MIU\textsuperscript{2}Net if we know the convergence power spectrum. We expect to reach the same $4\%$ uncertainty as in Sect.~\ref{sec:power-spectrum} assuming MIU\textsuperscript{2}Net is trained on the correct power spectrum. The input signal power spectrum can be estimated in the same way as for Wiener filtering without impacting the reconstruction result. 

For other statistics, Fig.~\ref{fig:cosmo2 rmse} shows that MIU\textsuperscript{2}Net's reconstruction RMSE under different smoothing conditions is almost unchanged, demonstrating that MIU\textsuperscript{2}Net reliably reconstructs the bulk characteristics of the convergence maps and maintains the same level of improvement over other methods. {\color{black} Additionally, we show in Fig.~\ref{fig:mass sheet} and Fig.~\ref{fig:mg_mass sheet} that the correction factor for reconstructing mean convergence generalizes very well across different cosmologies.} In the future, we aim to improve our simulation so that MIU\textsuperscript{2}Net is trained on a variety of cosmologies, so that the regularisation is not skewed to a single power spectrum. With our current implementation, however, MIU\textsuperscript{2}Net already generalizes reliably across cosmologies despite being trained on a single, arbitrary cosmology. 

\begin{figure*}
    \centering
    \includegraphics[width=0.8\linewidth]{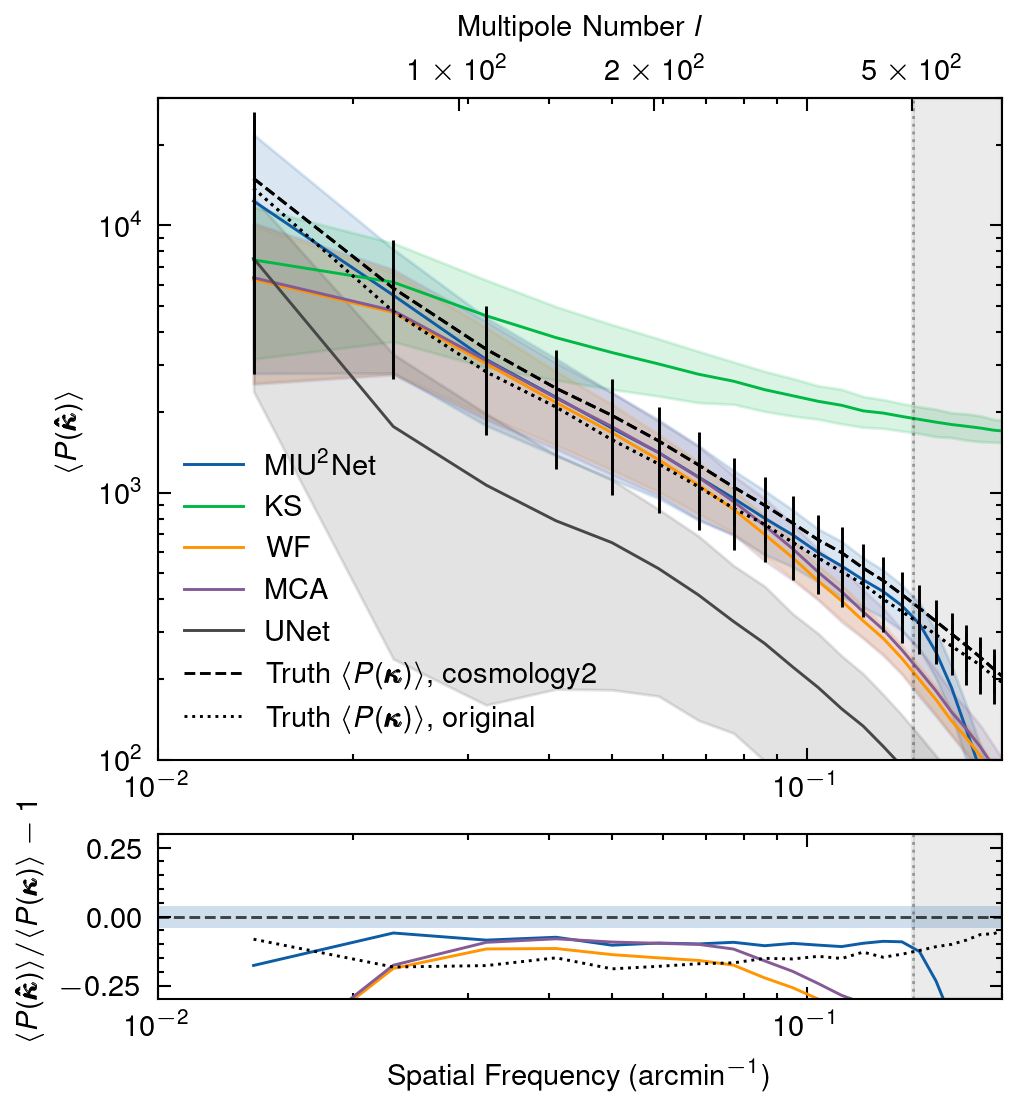}
    \caption{{\color{black}Power spectrum comparison for the second cosmology. \textit{Top panel:} radial-averaged power spectra for truth (new test cosmology), truth (original trained cosmology), MIU\textsuperscript{2}Net, KS, WF, MCA, and {\color{black} UNet}. The black dashed line and the error bars show the mean and the standard deviation for the true power spectrum for 500 realizations, under the new cosmology. The black dotted line shows the true power spectrum for the original power spectrum. Other coloured lines and shaded areas follow the same conventions in Fig.~\ref{fig:power spectrum}. There is a significant difference between the power spectra for the two cosmologies. \textit{Bottom panel:} residual power spectra for each reconstruction method with respect to the truth. The black dotted line shows the residual power spectrum of the original cosmology with respect to the new one. The blue band bounds the same region $[-0.04, 0.04]$ as in Fig.~\ref{fig:power spectrum}. Although trained on a different cosmology, MIU\textsuperscript{2}Net reaches the same performance as MCA (with the correct power spectrum as input) in intermediate spatial frequencies, and outperforms all methods in low and high spatial frequencies. Notably, MIU\textsuperscript{2}Net deviates from the trained dotted line and move closer to the true power spectrum.}}
    \label{fig:cosmo2 power spectrum}
\end{figure*}

\begin{figure*}
    \centering
    \includegraphics[width=0.6\linewidth]{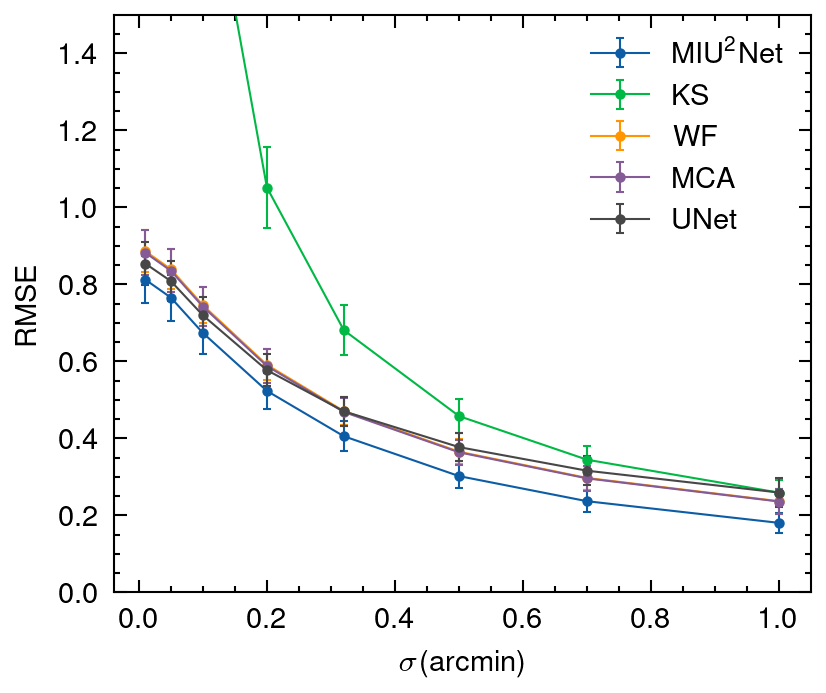}
    \caption{{\color{black}Root Mean Square Error (RMSE) under different smoothing conditions for the second cosmology. There is no significant difference between Fig.~\ref{fig:rmse} and RMSE with a different cosmology.}}
    \label{fig:cosmo2 rmse}
\end{figure*}

}

% =======================================================

\subsection{Full MIU\textsuperscript{2}Net Results}\label{sec:full-miu2net-results}

\begin{figure*}
    \centering
    \includegraphics[width=0.75\linewidth]{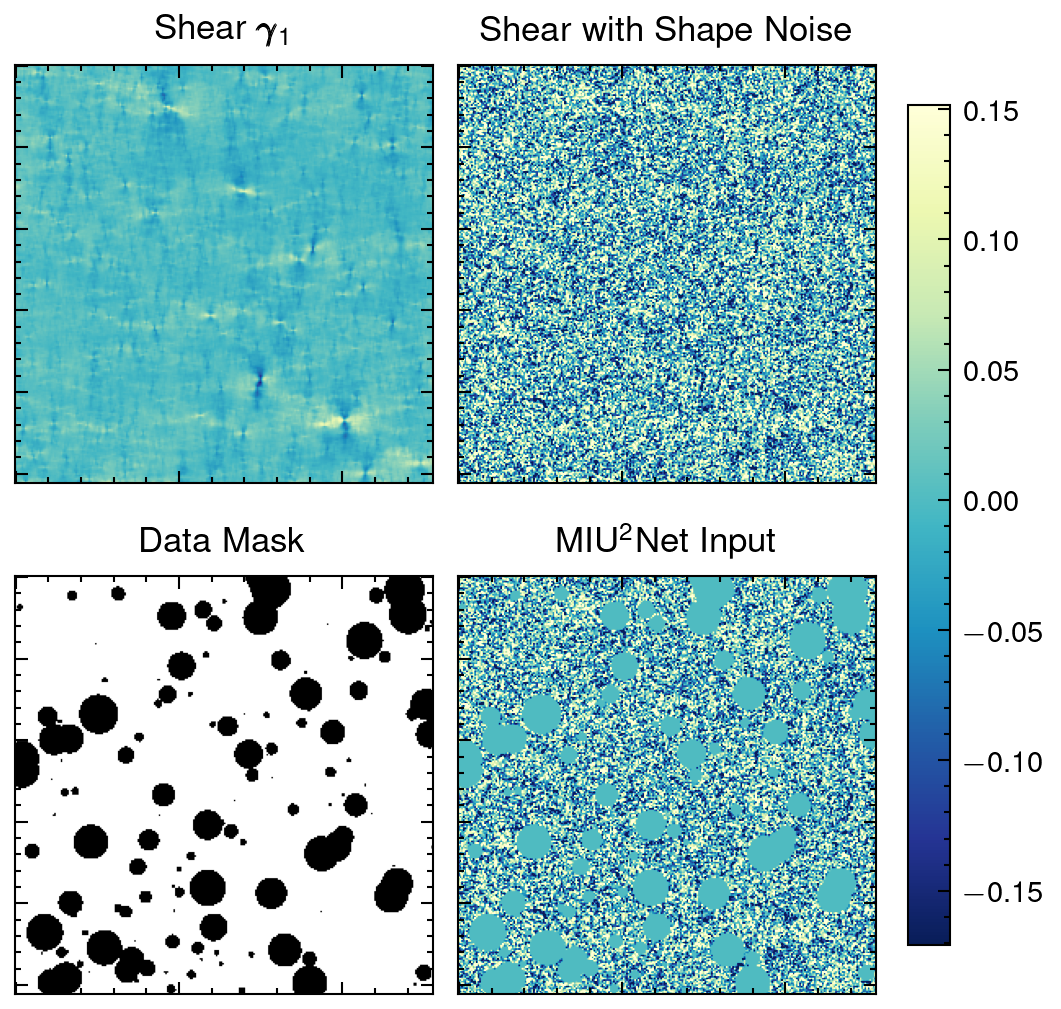}
    \caption{Visual comparison between shear, reduced shear, data mask, and MIU\textsuperscript{2}Net input data. \emph{Top left panel}: first component of a noiseless shear map \(\bm\gamma_1\). \emph{Top right panel}: {\color{black} the same first component of the shear field, with an added shape noise corresponding to \(n_g = 20 \, \rm{arcmin}^{-2}\).} \emph{Bottom left panel}: one mask realization covering \(\sim 20\%\) of the pixels in the map. {\color{black}Note that the coverage percentage is randomized during training.} \emph{Bottom right panel}: the shear field with data mask and shape noise corresponding to \(n_g = 20 \, \rm{arcmin}^{-2}\). The masked regions are set to \(0\). This is what the full MIU\textsuperscript{2}Net sees as its input (in the first channel; the second channel is the second component of noisy, masked reduced shear). We draw a new data mask realization and a shape noise realization for every training step.}
    \label{fig:shear}
\end{figure*}

Throughout the paper, we have only added shape noise to shear fields for clearer comparisons between MIU\textsuperscript{2}Net and established methods. {\color{black} However, data mask is another important observational constraint on shear that introduces significant systematics to mass inversion. Masks due to complicated survey geometry, bright stars, cosmic rays, and other observational effects can significantly impact the reconstruction quality. In this section, we include both shape noise and data masks on shear fields} and demonstrate MIU\textsuperscript{2}Net's ability to recover the correct summary statistics when both limitations are present. {\color{black} Importantly, we have not retrained our model, but only added the extra limitations during model inference.} We show a typical training data realization with both limitations in the bottom right panel of Fig.~\ref{fig:shear}. MIU\textsuperscript{2}Net can be further extended and customized by using survey-specific data masks, PSF uncertainties, and other systematic in the augmentation pipeline for training data.

{\color{black}Traditionally, different inpainting methods are used to fill the regions where data is missing. One popular inpainting method uses the Discrete Cosine Transform (DCT), as in \citet{Pires2009KS+, Pires2020KS+}. Because the convergence map \(\bm\kappa\) is sparse in DCT space, the strongest DCT coefficients encompass most of the signal, with smaller DCT coefficients describing the effects of noise and missing data. An iterative algorithm can then be used to estimate and enforce a convergence power spectrum in regions of missing data (e.g., \citealt{SchneiderSeitz1995-II, Jullo2014}). In this section, we test MIU\textsuperscript{2}Net ability on masked data by inferencing on shear maps with $\sim 20\%$ pixels set to $0$, as shown in the bottom left panel of Fig.~\ref{fig:shear}. The sizes and shapes of those masks roughly correspond to bright star distributions. MIU\textsuperscript{2}Net naturally synthesizes information from all map regions to estimate the power spectrum. Because of RAPS loss, the masked regions are automatically inpainted with values following the convergence power spectrum, together with highly sophisticated considerations for border and near-field pixels. MIU\textsuperscript{2}Net thus treats data masks in a similar vein as DCT, but simpler to implement and much quicker as it does not involve iterations once trained.}

{\color{black}The difference between shear and reduced shear is another important systematic in mass inversion. According to Eq.~\((\ref{linear})\), there is a linear relation between the shear field \(\gamma\) and the convergence field \(\kappa\). We have the observable in the form of reduced shear \(g = \gamma / (1 - \kappa)\) where \(g \simeq \gamma\) for most weak lensing regions. However, for small angular scales and regions near convergence peaks, \(\kappa\) is not negligible and the linear relation in \((\ref{linear})\) breaks down \citep{White2005SmallAngle, Shapiro2009SmallAngle}. Variants of traditional methods like KS+ \citep{Pires2020KS+} use an iterative scheme to estimate the convergence map. In the case of MIU\textsuperscript{2}Net, training with reduced shear instead of shear is theoretically beneficial. On the one hand, MIU\textsuperscript{2}Net does not assume a linear inverse problem but directly learns the mapping, so it is not restricted to a linear reconstruction as KS or WF. On the other hand, using the reduced shear \(g\) almost always leads to a higher SNR because \(\kappa\) is positive in most regions. For regions near convergence peaks, the boost in SNR is the most prominent, leading to better reconstruction for non-linear structures in the mass density field. In principle, the stronger the discrepancies between shear and reduced shear, the better MIU\textsuperscript{2}Net recovers non-Gaussian summary statistics. For fair comparisons, we have not included the reduced shear correction in training MIU\textsuperscript{2}Net. However, it is easy to incorporate reduced shear and further improve the reconstruction quality by augmenting the training set with equation (\ref{eq:reduced-shear}).}

We show the results of full MIU\textsuperscript{2}Net with {\color{black}shape noise and data masks} in Appendix~\ref{sec:additional-figures}. Figures~\ref{fig:mg_visual}--\ref{fig:mg_rmse} demonstrate that the MIU\textsuperscript{2}Net reconstructions almost do not degrade compared to MIU\textsuperscript{2}Net with only shape noise. Fig.~\ref{fig:mg_visual} and Fig.~\ref{fig:mg_profile3d} show that the reconstruction for peak centroids and amplitudes {\color{black}is accurate, despite the convergence peak about to fall into a masked area.} Fig.~\ref{fig:mg_dynamic range} illustrates that MIU\textsuperscript{2}Net reconstructions still recover the true dynamic range with minimal scatter. {\color{black}On the mass-sheet degeneracy, MIU\textsuperscript{2}Net reconstructions again reflect the correct mean convergence (Fig.~\ref{fig:mg_mass sheet}) with the same correction factor \(\mu_{\rm ms} = 1.672\), and remain robust under the second cosmology in the right panel of Fig.~\ref{fig:mg_mass sheet}.} The convergence KDE curves in Fig.~\ref{fig:mg_distribution} show MIU\textsuperscript{2}Net's ability to reproduce the (normalized) true convergence distribution even with \(\sim 20\%\) of masked data. We plot the power spectrum in Fig.~\ref{fig:mg_power spectrum} and show that MIU\textsuperscript{2}Net's constraint on the power spectrum {\color{black}is almost unchanged, achieving \(4\%\) uncertainties over all scales \(0 < l \lesssim 500\).} KS and WF, however, greatly suffer from missing data and the non-linear relationship between reduced shear and convergence. {\color{black}In Fig.~\ref{fig:mg_rmse}, we show that reconstruction RMSE under all limitations is not significantly different from the shape-noise-only RMSE in Fig.~\ref{fig:rmse}.} In all tests and summary statistics, MIU\textsuperscript{2}Net achieves comparable results with minimal degradation. This versatility against different forms of observational constraints far exceeds traditional methods where individual iterative algorithms have to be developed to treat each term deviating from the ideal assumption. More complicated dependencies (e.g. how reduced shear with PSF uncertainties influence convergence) that are difficult to solve inversely can now be learned by MIU\textsuperscript{2}Net by simply forward-modelling them in the training set. We can easily customize MIU\textsuperscript{2}Net by connecting the training set to mock galaxy catalogues and other survey-specific simulations as well.

\section{Conclusion}\label{sec:conclusion}

In this work, we develop MIU\textsuperscript{2}Net, a {\color{black}versatile} deep learning framework for weak lensing mass inversion. {\color{black}To regularise our reconstruction, }we construct a combined loss function to jointly estimate a pixel-wise convergence map and its frequency-domain power distribution, recovering the convergence power spectrum with \(4\%\) uncertainties up to very small angular scales \(l \simeq 500\). In addition to two-point statistics, MIU\textsuperscript{2}Net accurately recovers the {\color{black}convergence distribution, peak centroid, and peak amplitude.} The recovered convergence maps also follow the true log-normal distribution with an extended high-end tail. MIU\textsuperscript{2}Net is versatile against different forms of systematics such as shape noise, reduced shear, and data masks. Compared with {\color{black}unsupervised methods such as Kaiser-Squires, Wiener filtering, and MCALens, as well as supervised learning-based methods such as DeepMass,} MIU\textsuperscript{2}Net achieves superior reconstruction quality and better error handling in all metrics and summary statistics. {\color{black}MIU\textsuperscript{2}Net is also able to generalize across cosmologies despite being trained on a fixed cosmology.} 

In the era of Stage IV surveys, many large-scale weak lensing projects are joining forces, among which are the Chinese Space Station Telescope (CSST; \citealt{ZhanHu2011CSST, ZhanHu2021CSST, Yao2024csst}), \emph{Euclid} \citep{Euclid2011}, Vera Rubin Observatory's Legacy Survey of Space and Time (LSST; \citealt{LSST2019}), and the Nancy Grace Roman Space Telescope \citep{Roman2019}. With massive sky coverage and high galaxy density, these surveys will provide us with unprecedented weak lensing data. MIU\textsuperscript{2}Net is well-positioned to significantly contribute to the construction of mass maps from these future surveys. {\color{black}Although the error on power spectrum is yet insufficient to constrain cosmological parameters using convergence maps, we demonstrate that power spectrum regularisation is effective and our deep learning algorithm is versatile for accurate mapping of convergence distribution. MIU\textsuperscript{2}Net can open many possibilities, including dark matter halo identification, statistics, and studies on galaxy evolution under different dark matter environments.} By generating precise dark matter mass distributions, our method will not only become an essential tool for {\color{black}dark matter mass mapping} but also provide valuable insights into the interplay between the baryonic world and the dark matter universe.

% \section*{Acknowledgements}

\begin{acknowledgements}
We thank Mingxiang Fu for managing the GPU server, and Yufei Meng for useful discussions. The cosmological simulations were mainly conducted on the Yunnan University Astronomy Supercomputer. RL acknowledges the support by National Key R\&D Program of China No. 2022YFF0503403, the support of National Nature Science Foundation of China (Nos. 11988101), the support from the Ministry of Science and Technology of China (Nos. 2020SKA0110100),  the science research grants from the China Manned Space Project (Nos. CMS-CSST-2025-A03), CAS Project for Young Scientists in Basic Research (No. YSBR-062), and the support from K.C.Wong Education Foundation. XL acknowledges support from the NSFC of China under grants No. 12173033 and No. 11803028, National Key R\&D Program of China No. 2022YFF0503403, YNU grant No. C176220100008, and the research grants from the China Manned Space Project No. CMS-CSST-2021-B01.

All \verb|PYTHON| code and data that we used to generate the figures in the paper are available at \href{https://github.com/MIU2NET/miu2net}{https://github.com/MIU2NET/miu2net}. We have also included all $500$ convergence maps reconstructed independently by MIU\textsuperscript{2}Net, KS, and WF, which are used to generate the figures. The majority of KS and WF reconstruction code are adapted from the publicly available \verb|cosmostat| package at \href{https://github.com/CosmoStat/cosmostat}{https://github.com/CosmoStat/cosmostat}. 
\end{acknowledgements}

\bibliographystyle{raa} % style aa.bst
\bibliography{paper_bib} % your references Yourfile.bib

\appendix

\section{Training Specifications}\label{sec:training-spec}

We detail our training specifications and hyperparameters for MIU\textsuperscript{2}Net in Table~\ref{tab:training specs}. The full model takes \(\sim 1.5\) minutes per epoch to train on one NVIDIA A100 GPU, and the validation loss usually plateaus after \(\sim 256\) epochs. However, slightly better results may be achieved with greater batch size and more training epochs. Below, we show the hyperparameters used to train the specific model analyzed in this paper.

\begin{table*}
\caption{Training specifications for MIU$^2$Net.}
\label{tab:training specs}
\begin{tabular}{p{0.15\linewidth} p{0.1\linewidth} p{0.1\linewidth} p{0.5\linewidth}}
    \hline
    Name & Symbol & Value & Notes \\
    \hline
    epoch & e & \(2000\) & Total training epochs. Training usually plateaus after $\sim 256$ epochs and has very little improvement afterwards. \\
    batch size & b & \(128\) & Batch size used to train the final model. Any batch size greater than $16$ achieves comparable results. \\
    learning rate & lr & \(1\times10^{-4}\) & Initial learning rate. This
    learning rate is adjusted every epoch following a cosine annealing
    schedule with a final learning rate of \(1 \times 10^{-10}\). \\
    alpha & \(\alpha\) & \(1.0\) & Weight for MSE loss; see equation
    \((\ref{loss})\). \\
    beta & \(\beta\) & \(3.0\) & Weight for RAPS Loss; see equation
    \((\ref{loss})\). \\
    cutoff radius & \(r_{\max}\) & \(16.0\) & Cutoff radius for RAPS loss in
    units of pixel; see Eq.~\((\ref{raps})\). \\
    weight decay & wd & \(0\) & Weight decay for AdamW optimizer. \\
    delta & \(\delta\) & \(50.0\) & The threshold where Huber loss changes
    between MSE loss and $l1$ loss. During training, we use Huber loss instead
    of plain MSE loss for the MSE term \(l_{\rm MSE}\) to prevent gradient
    explosion in the very first few training steps. A very high delta
    ensures Huber loss to be identical to MSE loss once training stabilizes. \\
    galaxy number & \(n_g\) & \(20.0\) & Number of galaxies per square
    arcmin used to estimate shape noise. \\
    mask fraction & mf & {\color{black}\(0 - 25\%\)} & Fraction of pixels to be masked in an input shear or reduced shear map. {\color{black}Percentage randomized every training step.}\\
    \hline
\end{tabular}
\end{table*}

\section{Additional Figures}\label{sec:additional-figures}

\begin{figure*}
    \centering
    \includegraphics[width=1\linewidth]{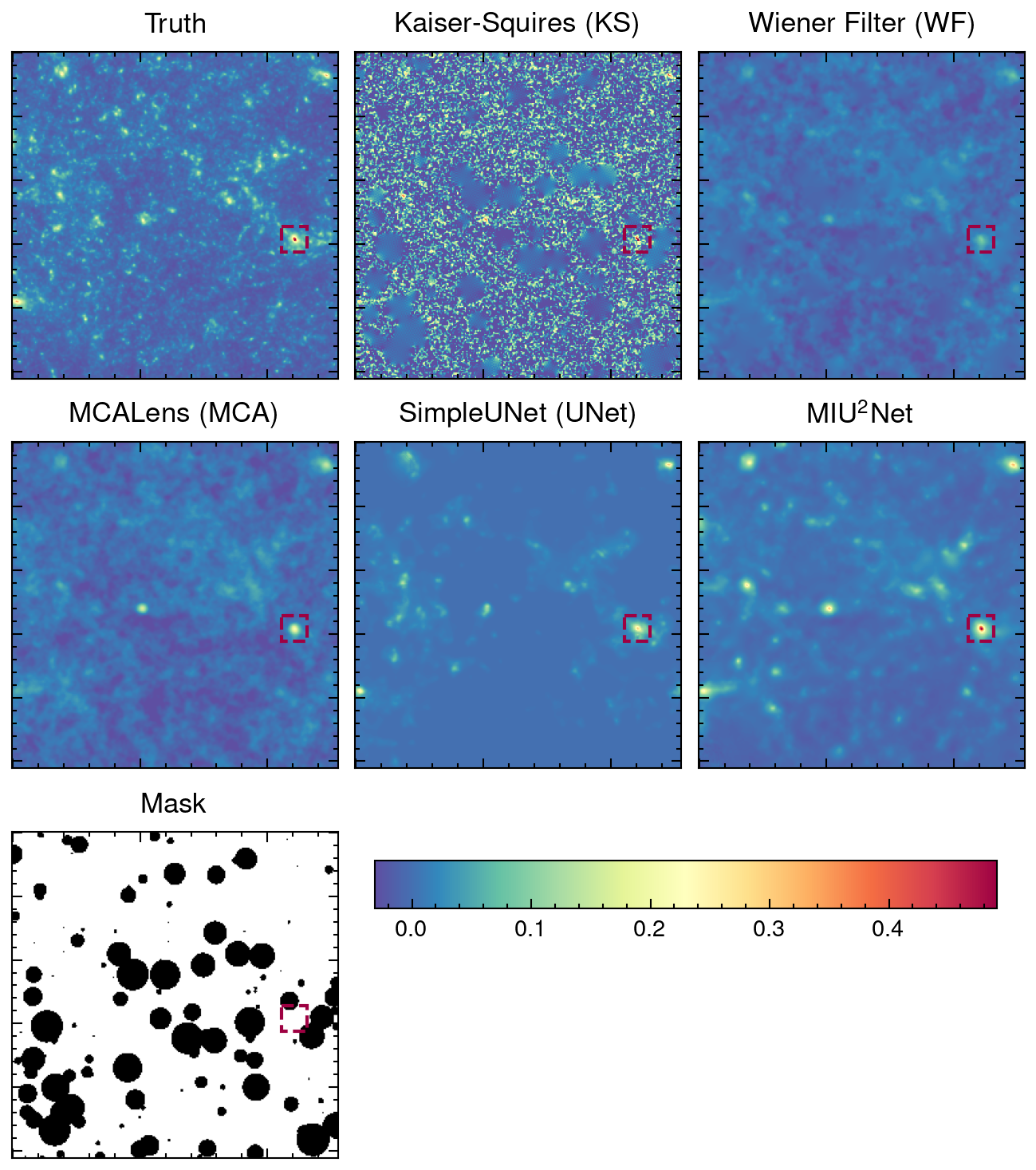}
    \caption{{\color{black}Visual comparison between Truth, KS, WF, MCA, {\color{black} UNet}, and MIU$^2$Net convergence map reconstructed from a masked noisy shear field.} The underlying noiseless shear is the same realization as in Fig.~\ref{fig:visual}. We can see the shape of the data mask in the blue-green circular regions in the KS reconstruction, because KS does not handle noise or data masks. We visualize in detail the scarlet dashed square region in each panel (the same regions as in Fig.~\ref{fig:visual}) in Fig.~\ref{fig:mg_profile3d}. {\color{black}Note that the central peak in the scarlet square falls right into a masked area.}}
    \label{fig:mg_visual}
\end{figure*}

\begin{figure*}
    \centering
    \includegraphics[width=0.75\linewidth]{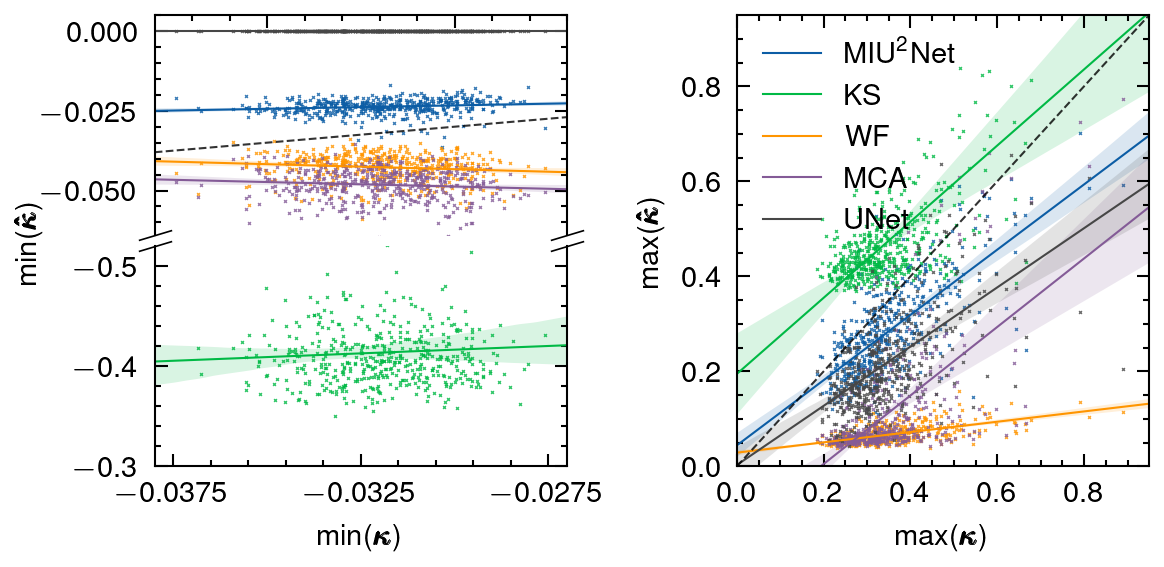}
    \caption{{\color{black}Dynamic range comparison for MIU$^2$Net, KS, WF, MCA, and {\color{black} UNet} under shape noise and data masks.} \textit{Left panel}: we plot the same minimum predicted convergence value $\min(\pmb{\hat\kappa})$ against minimum true convergence value $\min(\pmb\kappa)$ for each reconstruction, as in Fig.~\ref{fig:dynamic range}. {\color{black} We also include the best-fit line and $95\%$ confidence interval.} MIU$^2$Net accurately recovers the lower bound. \textit{Right panel}: {\color{black} scatter plot, best-fit line, and $95\%$ confidence interval for maximum values for each truth-prediction pair.} Despite the increased noise in input maps, the best-fit line for MIU$^2$Net is the closest to the ideal reconstruction of the upper bound.}
    % Although KS appears to have the right slope, its $95\%$ confidence interval is very wide, showing that the KS reconstruction is closer to a cluster than a credible linear relation.
    \label{fig:mg_dynamic range}
\end{figure*}

\begin{figure*}
    \centering
    \includegraphics[width=0.7\linewidth]{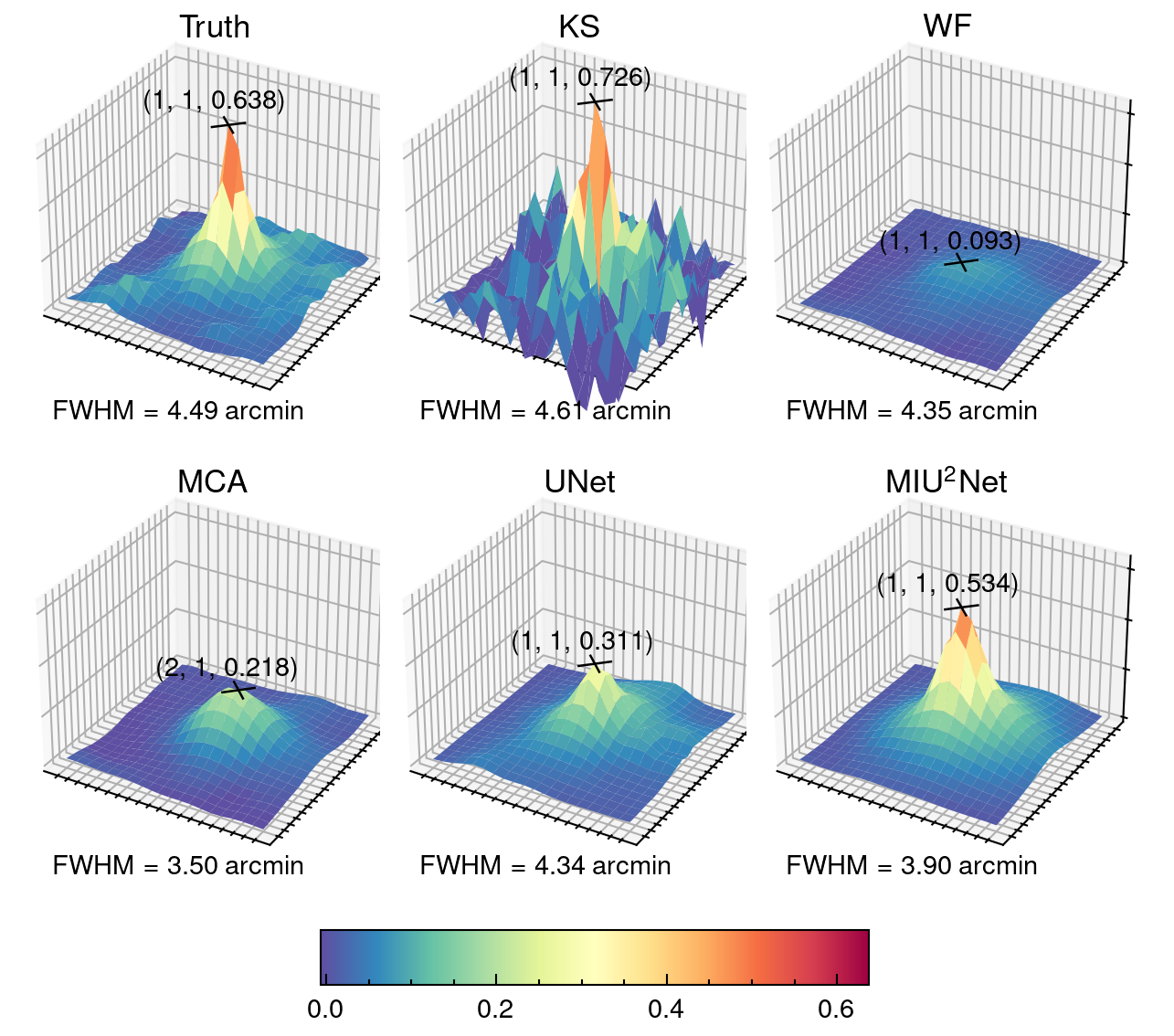}
    \caption{Parallel to Fig.~\ref{fig:profile3d}, we plot the centroid and amplitude for an individual convergence peak under shape noise and data masks. The plotted regions correspond to the scarlet dashed square regions in Fig.~\ref{fig:mg_visual}. {\color{black}Note that the true peak centroid is very close to a masked region in the shear map. Both MIU$^2$Net and {\color{black} UNet} recover a peak, but MIU$^2$Net has an amplitude closer to the truth.}}
    \label{fig:mg_profile3d}
\end{figure*}

\begin{figure*}
    \centering
    \includegraphics[width=1\linewidth]{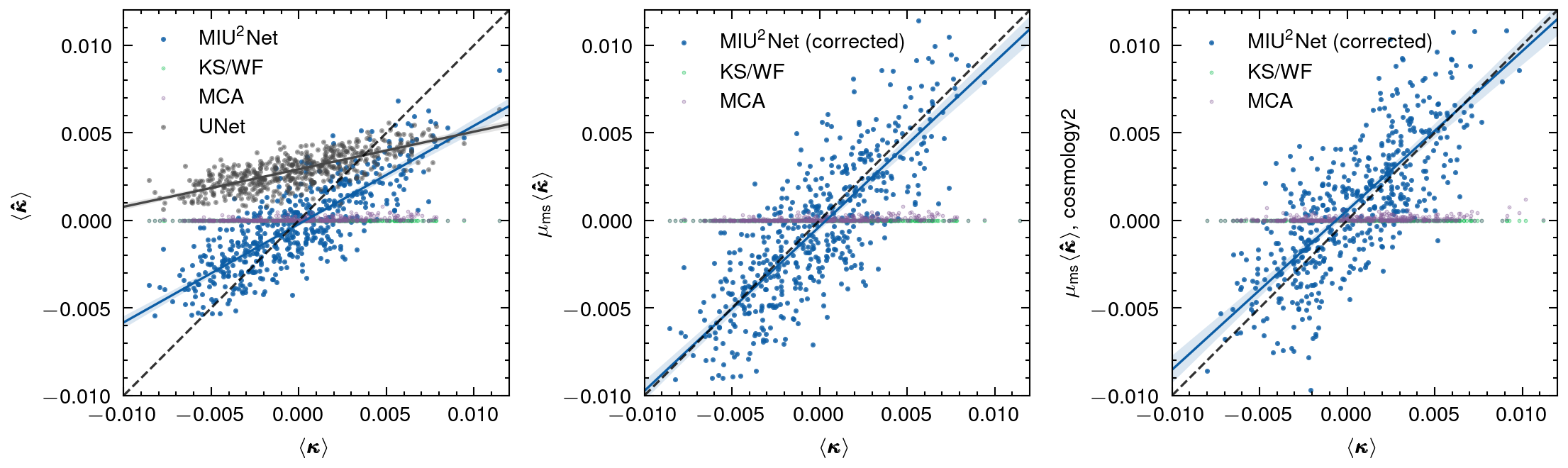}
    \caption{{\color{black}Mean convergence plot with shape noise and masked regions present in the reconstruction.} {\color{black}We use the same correction factor $\mu_{\rm ms} = 1.672$ as in Fig.~\ref{fig:mass sheet} in the middle panel. MIU$^2$Net reconstructions still follow the ideal linear relationship (black dashed line) with the correction factor based on $0\%$ masked fraction. In the right panel, we plot the mean convergence based on the second cosmology with both systematics.}}
    \label{fig:mg_mass sheet}
\end{figure*}

\begin{figure*}
    \centering
    \includegraphics[width=1\linewidth]{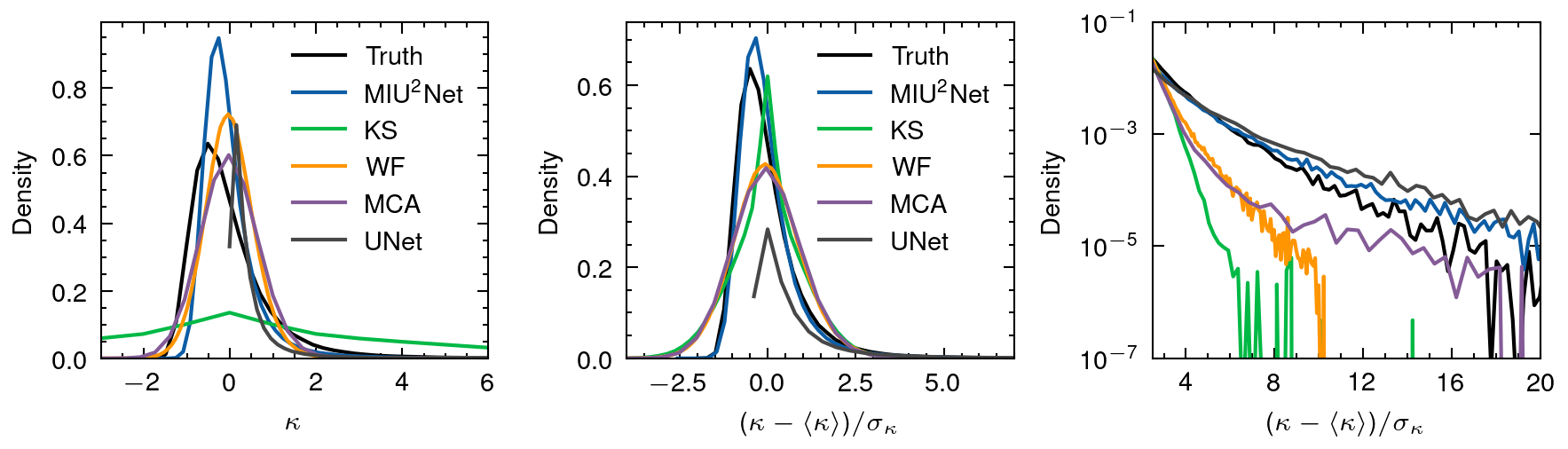}
    \caption{Kernel Density Estimation (KDE) showing the convergence distribution for all methods under shape noise and data masks covering $\sim 20\%$ of all pixels in the input maps. As shown in the middle panel and the right panel, MIU$^2$Net reproduces the true log-normal distribution with an extended high-end tail, whereas KS and WF significantly deviate from the true distribution. The shape of the MIU$^2$Net KDE curve is almost unchanged as in Fig.~\ref{fig:distribution}, demonstrating MIU$^2$Net's versatility against $\sim 20\%$ of missing data.}
    \label{fig:mg_distribution}
\end{figure*}

\begin{figure*}
    \centering
    \includegraphics[width=0.7\linewidth]{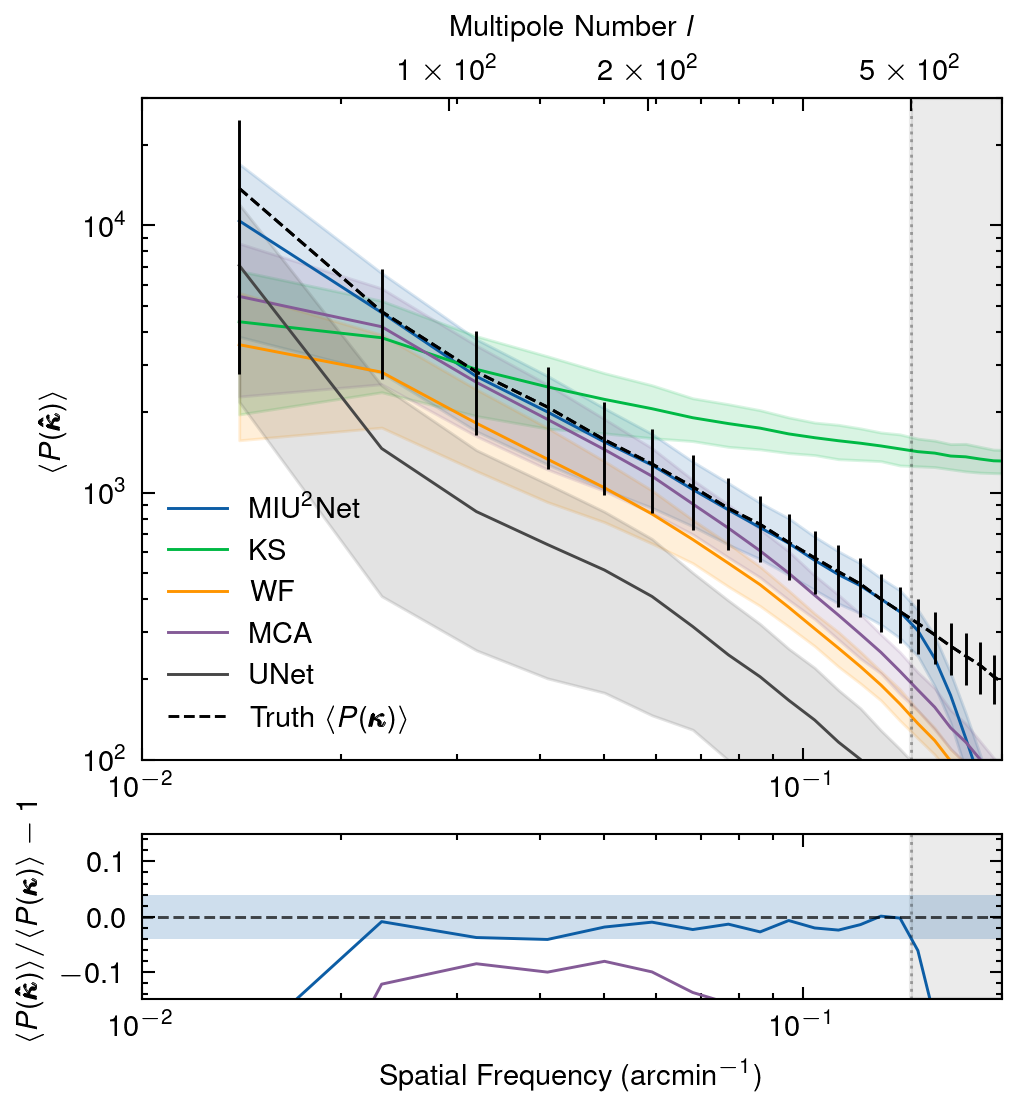}
    \caption{Radial-averaged power spectrum for each method under shape noise and data masks. {\color{black}Compared to Fig.~\ref{fig:power spectrum}, the MIU$^2$Net power spectrum is almost unchanged, achieving $4.0\%$ uncertainties over all scales $0 < l \lesssim 500$. In the bottom panel, all other methods are outside of the plot range except MCA.} The blue band in the bottom panel bounds the same $\pm 4\%$ uncertainty region as in Fig.~\ref{fig:power spectrum}. KS and WF power spectra degrade noticeably under missing data and the non-linear relationship between reduced shear and convergence. {\color{black}At $l \simeq 500$, MIU$^2$Net now achieves $1370\%$ improvement over WF (assuming $4\%$ error for MIU$^2$Net).}}
    \label{fig:mg_power spectrum}
\end{figure*}

\begin{figure*}
    \centering
    \includegraphics[width=0.6\linewidth]{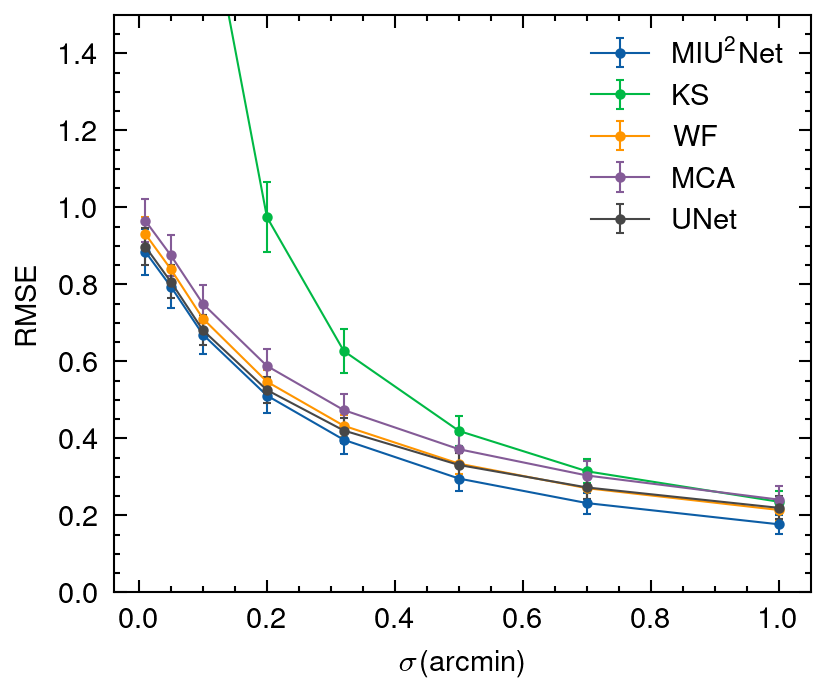}
    \caption{{\color{black}Root Mean Square Error (RMSE) under different smoothing conditions, under shape noise, reduced shear, and data mask. There is no significant difference between the shape-noise-only RMSE in Fig.~\ref{fig:rmse} and RMSE under all limitations.}}
    \label{fig:mg_rmse}
\end{figure*}

%-----------------------------------------------------------------

\end{document}